\documentclass{article}

\usepackage{arxiv}

\usepackage[utf8]{inputenc} 
\usepackage[T1]{fontenc}    
\usepackage{hyperref}       
\usepackage{url}            
\usepackage{booktabs}       
\usepackage{amsfonts}       
\usepackage{nicefrac}       
\usepackage{microtype}      
\usepackage{lipsum}
\usepackage{mathrsfs}  
\usepackage{graphicx}
\graphicspath{ {./images/} }
\usepackage{amsmath}
\usepackage{amssymb}
\usepackage{mathtools}
\usepackage[title]{appendix}
\usepackage{float}
\usepackage{natbib}
\usepackage{titlesec}
\usepackage{subcaption}
\usepackage{caption}
\usepackage{comment}
\usepackage{csquotes}
\setcitestyle{authoryear, open={(},close={)}}

\usepackage[title]{appendix}
\DeclareMathOperator*{\argmax}{argmax}

\title{Named Entity Swapping for Metadata Anonymization in a Text Corpus}
\numberwithin{equation}{section}

\author{
 Jan Greve\\
  Institute for Statistics and Mathematics\\
  WU Vienna\\
  Welthandelspl 1, 1020 Vienna\\
  \texttt{jgreve@wu.ac.at}
   \And
   Lukas Sablica \\
  Institute for Statistics and Mathematics\\
  WU Vienna\\
  Welthandelspl 1, 1020 Vienna\\
  \texttt{lsablica@wu.ac.at}
}

\begin{document}
\maketitle
\begin{abstract}
This work introduces an anonymization scheme for a corpus of texts to safeguard metadata from disclosure. It specifically aims to prevent large language models from identifying metadata associated with texts, thereby avoiding their influence on query responses. The core mechanism is called named entity swapping, a technique inspired by data swapping in statistical disclosure control. Our method randomly selects pairs of semantically similar substrings from different texts based on the similarity of their embedding vectors and interchanges some named entities between them. This prevents certain combinations of named entities from being uniquely associated with the metadata of individual texts. Our approach offers two key advantages. First, it enables users to determine the optimal level of anonymization that balances data utility and data risk through a calibration of several key decision variables. Second, it leverages text embeddings both to compute swapping weights and to assess data utility, enabling a high degree of flexibility and customization in the overall workflow. The effectiveness of the proposed method is demonstrated with an application that prevents the disclosure of company names in a cross-sectional dataset of earnings call transcripts. 
\end{abstract}


\section{Introduction}
The role that text data plays in predictive tasks in socioeconomic applications has increased significantly in recent years. For example, \cite{Gentzkow} covers a wide variety of socioeconomic applications based on text data using conventional machine learning and statistics. More recently, it is also getting increasingly popular to use transformer-based  \citep{vaswani2017attention} large language models (LLMs) for such tasks. The rationale is that LLMs are pre-trained on a vast amount of text data and hence ``aware'' of the statistical properties of these texts, which has predictive power when performing these functions. Works such as \cite{lopez2023can} in finance, \cite{argyle2023out} in political science and \cite{horton2023large} in economics are a few recent LLM-based applications in these areas. 

However, the flexibility of LLMs to incorporate a vast amount of textual information from pre-training could also result in the model being ``aware'' of information that it is not intended to use when performing predictive tasks. For example, for forecasting-related applications such as return predictions, risk assessments and sentiment analysis, the look-ahead bias is of particular concern. This bias is commonly introduced when training a model using historical text data for forecasting purposes. Specifically, when training, the model attempts to use the information about future events included in the text data it has been pre-trained on. For example, an LLM may use information such as the 737 MAX scandal when asked to evaluate the risks that Boeing faces using their 10-K filings from the year 2018. These concerns have already been voiced in works such as \cite{glasserman2023assessing}, \cite{lopez2023can}, \cite{halawi2024approaching} and \cite{sarkar2024lookahead}. Particularly in \cite{sarkar2024lookahead}, they showed empirically in the a risk analysis task using earnings call transcripts that responses from LLMs pretrained in historical texts do indeed contain look-ahead bias, and that prompting-based measures to prevent these biases (e.g., to specifically instruct the LLM to use information up to a given time) do not result in complete elimination of them.

The objective of our work is to demonstrate a text anonymization scheme that aids in mitigating such biases. We achieve this through replacing pieces of information in a text that can be used by LLMs to obtain its metadata with another ``similar'' information from a different text. By this substitution, the ties to the metadata that the original pieces of information had will now be severed as the replacement piece is sourced from another document with different metadata. In this way, the original metadata of the text cannot be used by the LLM to associate it with real events it may be aware of through pre-trained texts, which is the source of aforementioned biases such as the look-ahead bias. In other words, the text becomes generic information deprived of its identity-disclosing features (or mixed with two different identity-disclosing features). Texts that went through this procedure will have majority of their words intact and are thus likely to retain some statistical properties of their original form. Therefore, this approach offers distinct advantages over more information-suppressive anonymization techniques, such as summarization.

The aforementioned anonymization scheme we described is essentially data swapping adapted to text data. Data swapping refers to a well-established method in statistical disclosure control literature \citep{DALENIUS198273,FienbergSwapping} that aims to anonymize observations with categorical attributes through interchanging some of the attribute values between different observations. Our adaptation of this procedure to text data is achieved through the use of named entity recognition (refer to \citealp{JurafskyMartin}, for the definition). Specifically, substrings within each text are viewed as observations, and categorical attributes are named entities present in each substring. Hence, swapping takes place between a pair of substrings (from different texts) by interchanging some of their named entities that belong to the same category (e.g., interchanging names between two substrings). For this reason, we call our anonymization scheme \textit{named entity swapping}.

Substrings after receiving named entity swapping will evidently experience some distortions to their semantic meaning. Therefore, it is preferable to pair substrings that share similar semantic roles. By doing so, the swapping of named entities between them will likely result in a small change in their overall semantic meaning. For this purpose, we will introduce a weighting scheme that improves the selection of pairs of substrings that receive swapping. This is achieved through clustering these substrings on their embedding (a latent real-valued vector representation of texts learned with massive text data) space via a spherical clustering algorithm shown in \cite{Circlus} and implemented in the package \texttt{spheroids} \citep{spheroids}.

The goal of our proposed named entity swapping procedure is to strike the right balance between metadata disclosure risks, and the overall utility of swapped data as a source of textual information. In \cite{GomatamDecision2005}, they phrased this as a constrained optimization problem involving the data risk measure and the data utility measure. This approach is also adapted to suit our setting involving text data. The data risk measure employed in this work is based on the estimate of the number of substrings with a unique combination of named entities in the population, or so-called population uniques in the statistical disclosure literature. The rationale is that some of those population uniques may be used to identify the metadata of the text they belong to. Specifically, a suitably transformed version of the estimator proposed in \cite{hoshino2001applying} will be used. For the data utility measure, we re-purpose the mixture of spherical distributions which was used to generate swapping weights as an approximation of the true data-generating mechanism for the corpus of texts at hand. By doing so, we interpret the suitably transformed version of the likelihood ratio between the post-swapped text data and the original text data under this mixture model as a measure of change in data utility. The resulting constrained optimization problem between data risk and data utility is solved assuming a linear trade-off relationship in these two measures.

To demonstrate the utility of the proposed named entity swapping procedure, we apply this methodology to the anonymization of company names in earnings call transcript data. Specifically, we employ a cross-sectional sample of transcripts from multiple firms within a given industry sector and show empirically that the proposed named entity swapping scheme significantly reduces the predictability of company names for two state-of-the-art LLMs: GPT-4o and Gemini 2.5 Flash.

Overall, the main advantages of our approach are its customizability and transparency. In various parts of the named entity swapping scheme, text embeddings are used, which are latent vector representations of texts that capture semantic meaning and contextual relationships learned from large text corpora. Therefore, any recent advances in this area can be directly incorporated into our work for increased performance. For example, it can incorporate a recent development in generating domain-specific embeddings through INSTRUCTOR embedders \citep{su2022one} to make the anonymization scheme more tailored to the specific type of text data at hand. This flexibility to adapt to recent advances in natural language processing is coupled with the transparency in outputs from the algorithms and statistical tools we use to facilitate anonymization. Those are mixture models, species sampling distributions, and constrained optimization, all mature tools with various established methods to assess the validity of the outputs and convergence. Hence, it compares favorably to more LLM-reliant anonymization measures with reduced transparency in the inner workings of the algorithms.

In Section \ref{sec:DecThrNamedEntitySwapping}, we introduce our named entity swapping scheme and the accompanying decision problem we solve. Specifically, it introduces named entity swapping as an adaptation of data swapping to text data. The proposed named entity swapping is further enhanced using a weighting scheme based on the use of spherical clustering of texts via their embedding loadings. The extent to which the corpus of texts undergoes swapping is controlled by several decision variables that are subject to an optimization problem. This decision-theoretic formulation of named entity swapping is an adaptation of the prior work by \cite{GomatamDecision2005} that phrases it as a constrained optimization problem involving measures of data risk and utility. The data risk measure we use in this work is introduced as an adaptation of a similar measure used in \cite{hoshino2001applying}, while the data utility measure uses the outputs from the aforementioned spherical clustering algorithm. Section \ref{sec:Application} demonstrates the use of named entity swapping in an application involving the anonymization of company names for a cross-section of earnings call transcripts. In particular, an in-depth coverage of data preprocessing and sensitivity analysis on hyperparameter choice is given, followed by a display of anonymization results, which showcase a significant reduction in the predictability of company names for swapped texts. Finally, some outputs of the named entity swapping are shown to highlight some characteristics of this anonymization scheme. Section \ref{sec:DiscConcl} concludes with some discussions.
\section{Decision Theoretic Formulation of Named Entity Swapping}\label{sec:DecThrNamedEntitySwapping}
Consider a corpus of $m$ documents, of which each document $i$ can be partitioned into $n_i$ non-overlapping substrings through preprocessing steps which will be detailed in Section \ref{subsec:DataPreProcessing}. We denote the total number of substrings across documents as $ n=\sum_{i=1}^m n_i$. These substrings are typically called text chunks and we denote the sequence of $n$ chunks as $\textbf{y} = (y_1,\ldots,y_n)$ where the first $n_1$ entries $(y_1,\ldots,y_{n_1})$ are text chunks of the first document in corpus. The accompanying metadata of a corpus of $m$ documents with a total of $n$ chunks is denoted as a sequence of categorical variables $\textbf{c} = (c_1,\ldots,c_n)$ of length $n$ with $m$ unique values. For example, the first $n_1$ entries $(c_1,\ldots,c_{n_1})$ correspond to metadata categories of the first $n_1$ chunks $(y_1,\ldots,y_{n_1})$ and thus each of them is set to $\text{CG}_1$, the categorical representation of metadata for the first document. Therefore, $\textbf{c}$ takes the form
\begin{align*}
    \textbf{c} =  (\underbrace{\text{CG}_1,\ldots,\text{CG}_1}_{n_1},\text{CG}_2,\ldots,\underbrace{\text{CG}_m,\ldots,\text{CG}_m}_{n_m}).
\end{align*}
Our proposed swapping procedure interchanges some named entities present in chunks $y_i$ and $y_j$ where $i \neq j$ and $c_i \neq c_j$ for $i,j\in\{1,\ldots,n\}$ and produces post-swap chunks $\tilde{y}_i$ and $\tilde{y}_j$. This operation aims to make the resulting post-swap chunks $\tilde{y}_i$ and $\tilde{y}_j$ harder to associate with the corresponding metadata $c_i$ and $c_j$ than pre-swapped chunks $y_i$ and $y_j$, thereby anonymizing $y_i$ and $y_j$ with respect to their metadtata $c_i$ and $c_j$. 

Crucially, this text anonymization scheme we propose is based on an assumption about a document's potential cause of metadata disclosure. The assumption is that the disclosure occurs solely through a certain combination of named entities within some text chunks. For example, some names, dates and locations that appear in a chunk when combined may uniquely identify its source document. Note that this assumption rules out the possibility of disclosure through the unique style of speech and/or writing. However, for our purpose, this type of disclosure risk appears negligible. 

Based on this assumption, each text chunk is converted into a cell entry in a $p$-way $k_1\times \cdots \times k_p$ contingency table of categorical variables
\begin{align*}
     \textbf{z} = \{z_{i_1\ldots i_p}\in\{0,1,\ldots,n\}; i_j=1,\ldots,k_j; j = 1,\ldots,p\}, \qquad \sum_{i_1,\ldots,i_p} z_{i_1\ldots i_p} = n
\end{align*}
where each variable $j$ represents the named entity category and $k_j$ is the total number of named entities that belong to category $j$ present in \textbf{y}. Thus, the cell count $z_{i_1\ldots i_p}$ represents the number of text chunks in $\textbf{y}$ with first named entity category equal to $i_1$, second equal to $i_2$ and so on. Once the text data $\textbf{y}$ is converted into a $p$-way contingency table of categorical variables \textbf{z}, it has a suitable form to receive the named entity swapping, which we cover in the subsequent parts.

One last ingredient for named entity swapping is the embedding vector $\textbf{x} = (x_1,\ldots,x_n)$ generated from text chunks $\textbf{y}$ using any of the readily available text embedding models. As is customary for most embedding models, we assume embedding vector loadings of dimension $d$ for each chunk $x_i\in\mathbb{R}^d, i = 1,\ldots,n$ to be normalized to have norm 1.

We call these three converted forms of a corpus of text data at hand as pre-swap data, and denote them as
\begin{align*}
    \mathscr{D}_\text{pre} = \{\textbf{x},\textbf{y},\textbf{z}\},
\end{align*}
which will be the central subject of named entity swapping. Specifically, the post-swap version of the above trio $\tilde{\textbf{x}}$, $\tilde{\textbf{y}}$ and $\tilde{\textbf{z}}$ is evaluated with respect to measures of metadata disclosure risk and data utility. Then, we use the decision-theoretic framework of data swapping proposed by \cite{GomatamDecision2005} to determine the optimal post-swap trio $\hat{\textbf{x}}$, $\hat{\textbf{y}}$ and $\hat{\textbf{z}}$ that gives the best trade-off in these two measures, out of all possible $\{\tilde{\textbf{x}}$, $\tilde{\textbf{y}},\tilde{\textbf{z}}\}$ under consideration. This entire workflow constitutes our proposed metadata disclosure control procedure centered around named entity swapping.
\subsection{Named Entity Swapping}\label{subsec:NES}
Data swapping \citep{DALENIUS198273,FienbergSwapping} is one of the widely accepted practices in statistical disclosure control that swaps the categorical (or ordinal) attribute values between a pair of samples in a data set. For a swap to be valid, at least one of the unswapped attributes between a pair of samples must be different. For example, if the authors of this paper are included in a dataset that only consists of columns of height and gender, swapping our heights will simply result in re-indexing our rows since we are both males. Hence, this pairing does not constitute a valid swap. Therefore, the number of swappable pairs for data with $n$ samples is maximally $\binom{n}{2}$ but could be less depending on the choice of swapping attributes and whether the resulting pairs constitute valid swaps or not.

\cite{GomatamDecision2005} focused on a setting where swapping takes place between a pair of observations that are chosen uniformly at random without replacement. This is called a random swap and one controls the number of samples that receive this treatment through the swap rate parameter $r\in[0,1]$, which represents the number of swapped observations relative to the sample size $n$. They phrased the statistical disclosure control problem of tabular categorical data using random swaps in a decision-theoretic framework where the swap rate $r$ and the set of attributes to be swapped are chosen in an optimal manner.

In our context, the data set $\textbf{y}$ consists of $n$ text chunks. Hence, each chunk is first converted into a set of named entities present in the chunk, and through this process gets converted into a cell count in the aforementioned $p$-way $k_1\times\cdots\times k_p$ contingency table of named entity categories $\textbf{z}$. Details of this preprocessing will be covered in Section \ref{subsec:DataPreProcessing}. In this way, in principle, we have the same data structure as in \cite{GomatamDecision2005} and hence, their methodology can be directly applied. However, in addition, we make some adjustments and improvements to further refine the workflow and suit our specific application involving text data, which differs in characteristics from data with categorical entries, even though it may be converted as such.

\begin{figure}[h]
    \centering
    \includegraphics[width=0.9\linewidth]{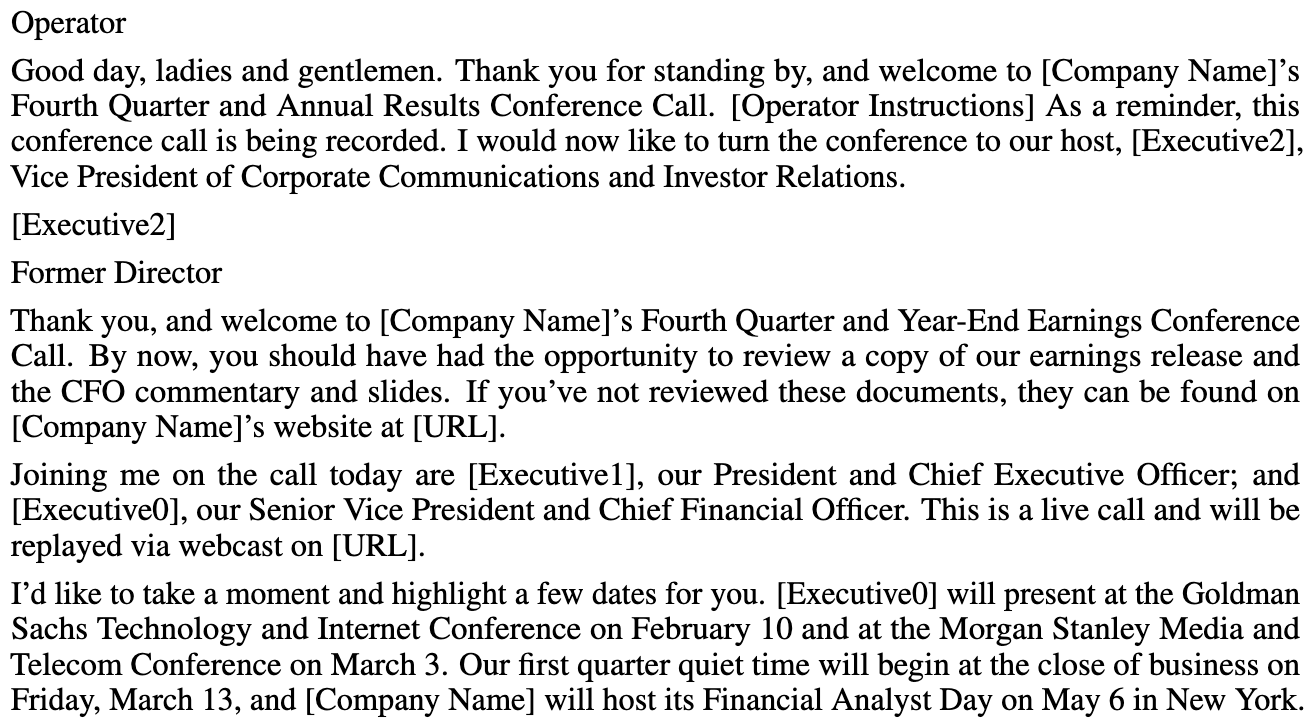}
    \caption{AMD Q4 2014 Earnings Call Opening with Direct Identifiers Suppressed}
    \label{fig:AMD_PostPreP}
\end{figure}

To understand the practical implications of swapping named entities in text data, consider an application that deals with the removal of pieces of information that disclose company names from earnings call transcripts. Figure \ref{fig:AMD_PostPreP} is an opening of an earnings call presentation preprocessed to have direct identifiers (information that on its own can be used to identify sensitive data, detailed in Section \ref{subsec:DataPreProcessing}) suppressed.

Specifically, it is an opening statement from the earnings call presentation of Advanced Micro Devices (AMD) in the fourth quarter of the year 2014, which took place on January 20th, 2015. Although most human analysts will likely struggle to identify the source of this text, many LLMs can do so. This is because of the terms at the very end of the last paragraph: ``Financial Analyst Day'', ``May 6'' and ``New York''. Pieces of information that act as direct identifiers when combined are called quasi-identifiers. In this case, these three terms, when combined, act as quasi-identifiers. Our proposed named entity swapping aims to break this type of potentially risky combination of named entities by interchanging some of these with entities that belong to the same category but differ in the entity names, which are sourced from a different document. In this example, we may consider ``Financial Analyst Day'' to belong to the category \texttt{Event},  ``May 6'' to \texttt{Date} and ``New York'' to \texttt{Location}. Hence, if one can find another chunk from a different company's earnings call transcript with \texttt{Date} and \texttt{Location} categories set to for instance ``Late March'' and  ``Texas'', and other categories including \texttt{Event} set to a value different from those of the AMD chunk (to constitute a valid swap), swapping may take place between these two chunks with respect to attributes \texttt{Date} and \texttt{Location}. The resulting post-swap AMD chunk now has \texttt{Event} = ``Financial Analyst Day'', \texttt{Date} = ``Late March'' and \texttt{Location}=``Texas'' which makes it difficult to predict that the text is from AMD earnings call even if the LLM somehow has the information about the event from pre-training data since the date and location are now different from actual ones.

The main advantage of swapping is that it keeps the majority of the statistics of the contingency table \textbf{z} unchanged. Firstly, it maintains the marginal distribution of any attributes (i.e., named entity categories). Secondly, it also maintains the joint distribution within both unswapped and swapped attributes. However, it does alter the joint distribution across unswapped and swapped attributes. In the above example, events like Financial Analyst Day usually take place around May or June in a location like New York, where the majority of the investment banks are headquartered. Therefore, swapping those attributes with an unlikely date and location, such as late March and Texas, does introduce a slight distortion to the data. To keep the level of distortion low, we introduce a weighting scheme to the random swapping procedure. This scheme is based on a spherical clustering algorithm that enables chunks with high semantic similarity to be more likely to be paired together for swapping.
\subsection{Clustering-Based Weighting Scheme}\label{subsec:ClustWeighting}
\subsubsection{Spherical Clustering of Embeddings of Earnings Call Text Chunks}\label{subsec:SphClstEmbdgs}
The introduction of weights in swapping has been proposed by  \cite{takemura1999local}. In \cite{GomatamDecision2005}, it is mentioned that their decision-theoretic framework of random swapping can be naturally extended to incorporate this approach. While \cite{takemura1999local} uses a matching-based weighting scheme, \cite{FienbergSwapping} mentions that clustering can be used for the same purpose.

We employ a weighting scheme based on a spherical clustering algorithm that clusters text chunks \textbf{y} through their embedding vectors \textbf{x}. Embeddings, or more specifically, text embeddings, are latent representations of the semantic roles of a text in the form of a $d$-dimensional real-valued vector of norm 1. As such, an embedding vector $x_i\in \textbf{x}$ generated from a text chunk $y_i \in \textbf{y}$ can be considered a point on a $d$-dimensional unit sphere $S^{d-1} =\{x\in\mathbb{R}^d:||x||=1\}$. Therefore, the closer a pair of embeddings $x_i$ and $x_j, i\neq j$ is in the spherical coordinates, the more similar the semantic meanings of their chunks $y_i$ and $y_j$ are. Hence, a cluster of embedding vectors is likely to contain text chunks that are semantically similar. This makes these chunks good candidates for named entity swapping, as named entities in those are likely to serve comparable roles in their respective texts, and thus may be close to being interchangeable.

To carry out the clustering of embedding vectors $\textbf{x}$ for the proposed weighting scheme, we use a spherical clustering algorithm based on Poisson kernel-based mixtures \citep{golzy2020poisson} and spherical Cauchy mixtures \cite{kato2020some} implemented in the \texttt{spheroids} package \citep{spheroids}. Our weighting scheme itself is simply uniform weights on all embeddings that belong to the same cluster and zero weights on embeddings that belong to all other clusters. Therefore, swapping will be performed uniformly at random between text chunks whose embeddings belong to the same cluster (provided that the chosen pair $(i,j)$ is from a different document and thus $c_i\neq c_j$ and other criteria for being a valid swap are met). Note that this approach can be further refined by, for example, incorporating the cosine similarity between embeddings in the same cluster.

The main advantage of this weighting scheme is its customizability. Firstly, the use of embeddings rather than term counts within each chunk makes this approach open to recent advances in the development of general-purpose and specialized text embeddings. In our work, we exclusively use OpenAI's ``text-embedding-3-small'' model \citep{openai2023embeddings} with $d = 1536$, but other more specialized embeddings trained on financial texts could also be considered for increased accuracy in the weighing scheme. Secondly, cluster-specific parameters we estimate using our approach could also be incorporated into the calculation of weights to further improve the pairing between semantically similar chunks. Finally, external covariates can also be included to control for additional information when performing clustering. For example, the metadata vector \textbf{c} can be incorporated to discourage chunks that belong to the same document from forming a cluster. In this way, one can incorporate known effects we would like to control for when searching for a matching pair.
\subsubsection{Poisson Kernel-Based and Spherical Cauchy Mixtures}
Embeddings generated from transformer-based architectures are typically normalized, as scale information in the vector entries is usually irrelevant. For this reason, distributions on spheres are the most natural candidates for a component distribution on mixture models for clustering embedding vectors $\textbf{x}=(x_1,\ldots,x_n), x_i\in S^{d-1},\forall i$. The most commonly used distribution for data on the unit sphere is the von Mises-Fisher (vMF) distribution \citep{khatri1977mises}. However, as this distribution belongs to the exponential family, the decay of the density function is exponential. This could result in too little variability in the clusters around the location parameter. For this reason, we use more flexible component distributions: the Poisson kernel-based (PKB) distribution \citep{golzy2020poisson} and the spherical Cauchy distribution \citep{kato2020some}.

A random vector on a $d$-dimensional unit sphere $x\in S^{d-1}$ that follows the PKB distribution has the following density:
\begin{align}
    f_\text{PKB}(x|\mu,\rho) = \frac{1-\rho^2}{||x-\rho\mu||^d},\qquad x\in S^{d-1}\label{eq:fPKB}
\end{align}
where $\mu\in S^{d-1}$ is the location parameter and the concentration parameter is $\rho\in[0,1)$. When $\rho = 0$, the distribution turns into a uniform distribution on the sphere, while $\rho\to 1^-$ makes the distribution converge to a Dirac mass at $x=\mu$. The spherical Cauchy distribution has a similar density:
\begin{align}
    f_\text{sCauchy}(x|\mu,\rho) = \Bigg(\frac{1-\rho^2}{||x-\rho\mu||^2}\Bigg)^{d-1},\qquad x\in S^{d-1}\label{eq:fsCauchy}
\end{align}
which coincides with the PKB density \eqref{eq:fPKB} when $d=2$. The role and the interpretation of location and concentration parameters $\mu\in\S^{d-1}$ and $\rho\in[0,1)$ are the same as in PKB distribution. The spherical Cauchy distribution has slightly lighter tails than the PKB distribution when $d>2$. Thus, it serves as a middle ground between the vMF distribution and the PKB distribution. Refer to \cite{Circlus} for more details regarding these component distributions. 
\subsubsection{Maximum Likelihood Estimation of PKB \& Spherical Cauchy Mixtures}
The numerical value of the MLE estimates of the PKB and spherical Cauchy mixtures are obtained using the EM-algorithm \citep{dempster1977maximum}. Specifically, for $n$ embeddings
\begin{align*}
    \textbf{x} = (x_1,\ldots,x_n), \qquad x_i\in S^{d-1},\qquad i=1,\ldots,n
\end{align*}
obtained from converting $n$ text chunks $\textbf{y} = (y_1,\ldots,y_n)$, the likelihood of a spherical mixture model with $K$ number of clusters is given as:
\begin{align}
f(x_1,\ldots,x_n|\pi_1,\ldots,\pi_K,\phi_1,\ldots,\phi_K) = \prod_{i=1}^n\sum_{k=1}^K \pi_kf(x_i|\phi_k)\label{eq:MixLikelihood}
\end{align}
for a density $f$, which is either PKB or spherical Cauchy given in \eqref{eq:fPKB} and \eqref{eq:fsCauchy}. The vector $(\pi_1,\ldots,\pi_K)$ is mixture component weights which sum up to one and $\phi_k$ are component-specific parameters which for these two densities are location $\mu_k$ and concentration $\rho_k$, thus $\phi_k= \{\mu_k,\rho_k\}$ for $k=1,\ldots, K$.

We use the EM algorithm to obtain MLE estimates of the parameters $(\pi_1,\ldots,\pi_K)$ and $(\phi_1,\ldots,\phi_K)$. Famously, the likelihood for location-scale family mixture distributions such as Gaussian mixtures are known to be unbounded, and an approach based on penalized maximum likelihood estimation is considered preferable \citep{chen2008inference}. This unboundedness also holds true for PKB and spherical Cauchy mixtures. One way to understand it is that if there are two points $x_1$ and $x_2$ in the same cluster, splitting them into two distinct clusters with sufficiently small concentration parameters will only increase the likelihood. However, it has been shown empirically in \cite{chen2009inference} that the local maxima of the unpenalized likelihood using the EM algorithm can still produce competitive estimates, provided that sufficient care for ruling out poor local maxima has been given. Moreover, the sole purpose of using a clustering algorithm for this application is to find candidate pairs of text chunks for swapping and not to obtain the ``true'' clustering solution. Therefore, even though the aforementioned cluster splitting will reduce the number of candidate pairs, the resulting smaller clusters will still contain chunks with similar embeddings. For the details of the EM-algorithm on spherical mixture models, we refer to \cite{Circlus} and the package \texttt{spheroids} \citep{spheroids} we use for this work. 

The clustering algorithm implemented in \texttt{spheroids} requires the user to specify two hyperparameters: the component density $f$, and the total number of clusters $K$. In addition, we specify another optional hyperparameter $\epsilon$, the lower bound for the mixture component weights $\pi_i, i= 1,\ldots,K$, to further control the clustering outcomes. In Section \ref{subsec:HyperParamChoice}, we inspect how the choice of these hyperparameters affects the resulting conclusions of the decision problem to determine the optimal level of swapping.
\subsection{The Optimization Problem and Decision Variables}\label{subsec:OptimizationProblem}
Our adaptation of the decision-theoretic formulation of data swapping introduced in \cite{GomatamDecision2005} has the following set of decision variables called releases $\mathcal{R}$:
\begin{align}
    \mathcal{R} = (r,NEC_1,\ldots,NEC_p,\mathcal{C})
\end{align}
where $r\in[0,1]$ is the aforementioned swap rate, $( NEC_1,\ldots,NEC_p)$ are named entity categories considered for swapping and $\mathcal{C}$ is a set of constraints. Each named entity category $NEC_i, i = 1,\ldots,p$ can take four values: swapped ($NEC_i = S$), must remain fixed ($NEC_i = F$), must change ($NEC_i = C$) and neither of those ($NEC_i = U$). For example, if a named entity category is deemed to be unsuitable for swapping $S$, but may (not must as in $C$) receive other forms of information suppression, it takes the value $U$. The constraints $\mathcal{C}$ we impose are two-fold. Firstly, the swapping of named entities can only be conducted between chunks that belong to different documents. That is, between $y_i$ and $y_j$ with $c_i\neq c_j$, $i,j\in\{1,\ldots,n\}$. This constraint reflects the nature of our application, where the metadata we protect from disclosure is defined at the level of the document and not at the level of each chunk. Secondly, the swapping takes place only between chunks that belong to the same cluster in the embedding space. As covered in Section \ref{subsec:SphClstEmbdgs}, we introduce a clustering-based weighting scheme to random swapping. The specific weighting scheme we propose is uniform weights on all chunks that belong to the same cluster and zero otherwise. Hence, this weighting scheme is incorporated as one of the constraints $\mathcal{C}$. In addition, there may be some domain-specific constraints or other constraints to reduce the computational burden of the optimization problem.

The release space $\mathscr{R}$ is the set of all releases $\mathcal{R}$. The cardinality of this set is combinatorial with respect to the number of named entity categories $p$ and thus, we work on a much smaller candidate release space $\mathscr{R}_\text{cand}$. For example, the candidate release space $\mathscr{R}_\text{cand}$ could be swapping of all two or three out of $p$ named entity categories for some select values of $r$. For each release $\mathcal{R}$ in candidate releases $\mathscr{R}_\text{cand}$, one can construct an actual release, which is the post-swap data $\mathscr{D}_\text{post}(\mathcal{R}) = \{\tilde{\textbf{x}},\tilde{\textbf{y}},\tilde{\textbf{z}}\}$ for suitably swapped $\tilde{\textbf{y}}$ and $\tilde{\textbf{z}}$ by choosing $\mathcal{R}$ and the embeddings $\tilde{\textbf{x}}$ generated by such $\tilde{\textbf{y}}$. The goal is to pick an optimal post-swap data $\mathscr{D}_\text{post}(\hat{\mathcal{R}}) = \{\hat{\textbf{x}},\hat{\textbf{y}},\hat{\textbf{z}}\}$ via choosing an optimal $\hat{\mathcal{R}}$ that gives the best trade-off in the reduction in disclosure risk and the loss of data utility compared to the pre-swap data $\mathscr{D}_\text{pre} = \{\textbf{x},\textbf{y},\textbf{z}\}$. 

One limitation of this decision theoretic framework proposed by \cite{GomatamDecision2005} is that the mapping from $\mathcal{R}$ to $\mathscr{D}_\text{post}(\mathcal{R})$ is not deterministic due to the randomness in the selection of swapping pairs. Therefore, from the actual candidate release space defined as:
\begin{align}
    \mathscr{R}_\text{cand}^\text{act} = \{\mathscr{D}_\text{post}(\mathcal{R}): \mathcal{R}\in\mathscr{R}_\text{cand}\}\nonumber,
\end{align}
it is possible to find a post-swap data $\hat{\mathscr{D}}_\text{post}(\mathcal{R})\in\mathscr{R}_\text{cand}^\text{act}$ from a suboptimal release $\mathcal{R}$ with better trade off in risk and utility than the ``optimal'' post-swap data $\mathscr{D}_\text{post}(\hat{\mathcal{R}}) \in\mathscr{R}_\text{cand}^\text{act}$ based on the optimal release $\hat{\mathcal{R}}$. To circumvent this issue, as briefly mentioned in \cite{GomatamDecision2005}, one may include the seed as part of the decision variable in $\mathcal{R}$ to be optimized. We instead define the expected post-swap data which is a deterministic function of the release $\mathcal{R}$ as follows
\begin{align*}
    \bar{\mathscr{D}}_\text{post}(\mathcal{R}) := \mathbb{E}(\mathscr{D}_\text{post}(\mathcal{R})).
\end{align*}
Then, we approximate this expectation through Monte Carlo integration
\begin{align}
    \bar{\mathscr{D}}_\text{post}(\mathcal{R}) \approx \frac{1}{M}\sum_{i=1}^M \mathscr{D}^{(i)}_\text{post}(\mathcal{R})\label{eq:MCintegration}
\end{align}
with $M$ random draws of $\mathscr{D}^{(i)}_\text{post}(\mathcal{R})$ given a release $\mathcal{R}$. Note however, that in our setting, the randomness in the mapping from $\mathcal{R}$ to $\mathscr{D}_\text{post}(\mathcal{R})$ is not as severe as in their original work. This is because the weighting scheme covered in Section \ref{subsec:ClustWeighting} drastically reduces the number of all possible swappings that can take place. Therefore, for the subsequent part we simply fix $M=1$ and claim that $\mathscr{D}_\text{post}(\mathcal{R})$ is a (crude) approximation of $\bar{\mathscr{D}}_\text{post}(\mathcal{R})$, a deterministic function of release $\mathcal{R}$. This is to give mathematical justification to the claim in \cite{GomatamDecision2005} to treat $\mathcal{R}\in\mathscr{R}_\text{cand}$ and $\mathscr{D}_\text{post}(\mathcal{R})\in\mathscr{R}_\text{cand}^\text{act}$ as equivalent in the subsequent optimization problem to pick optimal release $\hat{\mathcal{R}}$. In our empirical example introduced in Section \ref{subsec:EmpRslt}, we use this crude approximation to choose $\hat{\mathcal{R}}$ but also show predictive checks using sampled $\mathscr{D}^{(i)}_\text{post}(\mathcal{R})$'s to show that the variability in mapping from $\mathcal{R}$ to $\mathscr{D}_\text{post}(\mathcal{R})$ is not particularly large for our application. For applications where this is not the case, the following optimization problem involving data utility and data risk measures should be conducted with their Monte Carlo estimates with a large enough $M$.

From each post-swap data $\mathscr{D}_\text{post}(\mathcal{R})$, we compute data utility measure $DU:\mathscr{D}_\text{post}(\mathcal{R})\to[0,1]$ and data risk measure $DR:\mathscr{D}_\text{post}(\mathcal{R})\to[0,1]$. These measures are normalized to be relative to the base risk and base utility of the pre-swap data $\mathscr{D}_\text{pre}$. That is, when the swap rate $r$ is set to 0, it holds that $\mathscr{D}_\text{post}(\mathcal{R}) =\mathscr{D}_\text{pre}$, and $DU$ and $DR$ evaluated with respect to such $\mathscr{D}_\text{post}(\mathcal{R})$ are exactly 1. Therefore, as more swaps take place, the lower both measures become.

The decision problem to pick the optimal release $\hat{\mathcal{R}}$ under these measures is written as follows
\begin{align}
    \hat{\mathcal{R}} &= \argmax_{\mathcal{R}\in\mathscr{R}_\text{cand}} 
 \ DU(\mathscr{D}_\text{post}(\mathcal{R}))\label{eq:RUmaximization}\\
    &\text{s.t. } DR(\mathscr{D}_\text{post}(\mathcal{R}))\leq\alpha\nonumber.
\end{align}
which is the same as the optimization problem introduced in \cite{GomatamDecision2005}, with a superficial difference in the arguments given to $DU$ and $DR$ due to reasons mentioned above regarding their treatment of viewing $\mathcal{R}\in\mathscr{R}_\text{cand}$ and $\mathscr{D}_\text{post}(\mathcal{R})\in\mathscr{R}_\text{cand}^\text{act}$ as equivalent.

Another more general optimization problem they introduce, which works for small $\mathscr{R}_\text{cand}$ can be written as follows. Consider a partial order $\preceq_{\text{RU}}$ defined as 
\begin{align}
    \mathcal{R}_1\preceq_{\text{RU}}\mathcal{R}_2 \iff DR(\mathscr{D}_\text{post}(\mathcal{R}_2))\leq DR(\mathscr{D}_\text{post}(\mathcal{R}_1)) \qquad \text{and} \qquad DU(\mathscr{D}_\text{post}(\mathcal{R}_2))\geq DU(\mathscr{D}_\text{post}(\mathcal{R}_1)).
\end{align}
One can use this partial order to find maximal elements of $\mathscr{R}_\text{cand}$ and construct a risk-utility frontier $\partial \mathscr{R}_\text{cand}$ with respect to $\preceq_{\text{RU}}$. Then, assuming a linear risk-utility tradeoff of the form
\begin{align}
    DR =  c+ aDU,\label{eq:RU_LinEq}
\end{align}
which states that data risk $DR$ is a linear function of data utility $DU$ with the trade-off coefficient $a$ with an offset $c$. The latter term is introduced for added flexibility, as forcing the equation to go through the origin will limit the choice of the optimal $\hat{\mathcal{R}}$. As texts that receive more swapping with a lower risk will also have more distortion and thereby lower utility, the reasonable choice for the sign of the coefficient $a$ is positive. Once the user determines their right amount of $a$ and $c$, the release on the $\partial \mathscr{R}_\text{cand}$ that is tangent to the line \eqref{eq:RU_LinEq} is the optimal release $\hat{\mathcal{R}}$. In Section \ref{subseq:PredictiveChecks} we use this approach to identify the optimal $\hat{\mathcal{R}}$.
\subsection{Risk Measure}\label{subsec:RiskMeasure}
\subsubsection{Characteristics of the Risk Measure}
The disclosure risk measure $DR$ is a function on release space $\mathscr{R}$ of which we only consider the subset, the candidate release space $\mathscr{R}_\text{cand}$. More precisely, it is a function of the post-swap data $\mathscr{D}_\text{post}(\mathcal{R})\in\mathscr{R}_\text{cand}^\text{act}$ which from the aforementioned reasoning can be considered a crude approximation to $\bar{\mathscr{D}}_\text{post}(\mathcal{R})$, a deterministic function on $\mathcal{R}$. Hence we treat optimizing $DR$ relative to $DU$ with respect to $\mathcal{R}\in\mathscr{R}_\text{cand}$ and $\mathscr{D}_\text{post}(\mathcal{R})\in\mathscr{R}_\text{cand}^\text{act}$ as equivalent as in \cite{GomatamDecision2005}.

The risk measure $DR$ specifically depends on the $p$-way $k_1\times\cdots\times k_p$ contingency table $\tilde{\textbf{z}}$ in the post-swap data $\mathscr{D}_\text{post}(\mathcal{R}) = \{\tilde{\textbf{x}},\tilde{\textbf{y}},\tilde{\textbf{z}}\}$. We normalize the risk measure $DR$ to be 1 when evaluated on the contingency table of the pre-swap data $\textbf{z} \in \mathscr{D}_\text{pre}$ (which is post-swap data $\mathscr{D}_\text{post}(\mathcal{R})$ if $r=0$ so $DR$ is defined) and to be bounded between 0 and 1 for any post-swap $\tilde{\textbf{z}}\in  \mathscr{D}_\text{post}(\mathcal{R})$ with $r>0$. This allows us to interpret the risk measure $DR(\mathcal{R})$ as the ratio of post-swap risk relative to the base risk from the pre-swap data $\mathscr{D}_\text{pre}$.  

For each release $\mathcal{R} \in \mathscr{R}_\text{cand}$, we denote the set of named entity categories to be swapped as $J\subset \{NEC_1,\ldots,NEC_p\}, J\neq \emptyset$ and its cardinality as $|J|$. As we assume that identification risk translates to a combination of named entities acting as quasi-identifiers, we consider combinations of named entities in categories $J$ that are unique in the $p$-way contingency table \textbf{z} to be potentially risky. In other words, from the following marginalized $|J|$-way contingency table
\begin{align}
    \textbf{z}_J = \sum_{\substack{i_1,\ldots,i_p \\\text{except rows in } J}} z_{i_1\ldots i_p},
\end{align}
cells with one observation are considered risky. These observations are often called sample uniques in the statistical disclosure control literature. Note that cells that contain multiple chunks from the same company's earnings call also face disclosure risk. However, both $\textbf{z}$ and $\textbf{z}_J$ tend to be very sparse and hence such cells are usually rare. Therefore, we focus only on sample uniques when constructing a risk measure $DR$.

Rather than using the prevalence of sample uniques and other low-frequency counts in the contingency table as in \cite{GomatamDecision2005}, we assume a $|J|$-way marginalized superpopulation table $\textbf{Z}_J$ of size $N$ that generated the sample table $\textbf{z}_J$ of size $n$. Similar to its sample counterpart $\textbf{z}_J$, the table $\textbf{Z}_J$ is a $|J|$-dimensional marginal of the population table $\textbf{Z}$. The cells of the population table $\textbf{Z}$ represent all combinations of vocabularies of these named entity categories $NEC_1,\ldots,NEC_p$ in the corpus and not only limited to combination of named terms that are present in the corpus. $N$ is the superpopulation size that filled this table $\textbf{Z}$. Estimation of vocabulary and its frequency distribution is a topic commonly dealt with in superpopulation models (e.g., the work by \citealp{efron1976estimating}) and has also seen usage in statistical disclosure control literature for data without any textual context, such as \cite{samuels1998bayesian} and \cite{hoshino2001applying} as well. Hence, it is natural to consider a model in this area for our setting that deals with statistical disclosure control in textual data.

%
\subsubsection{Ewens-Pitman's Sampling Formula for Cell Frequency Counts}
To obtain an estimate of the number of population uniques in $\textbf{Z}_J$ using sample uniques in $\textbf{z}_J$, we follow the approach employed in \cite{samuels1998bayesian} and \cite{hoshino2001applying}. These works fit a class of sampling-consistent distributions on partition sizes of data, denoted as
\begin{align*}
    (s_1,\ldots,s_n), \qquad \sum_{j=1}^ns_j = k, \qquad \sum_{j=1}^njs_j = n.
\end{align*}
In our setting, partition sizes correspond to frequency counts of cells in the $|J|$-way contingency table $\textbf{z}_J$. That is
\begin{align*}
    s_j = \sum_{j_1,\ldots,j_{|J|}}1\{\textbf{z}_{j_1\ldots j_{|J|}} = j\}
\end{align*}
for all $|J|$ rows with cells $(j_1,\ldots,j_{|J|})\in J$. This approach posits the existence of a $p$-way population contingency table $\textbf{Z}$ of $N$ observations whose $|J|$-dimensional marginals $\textbf{Z}_J$ with respect to the set of named entities $J$ has nonzero frequency counts
\begin{align*}
    (S_1,\ldots,S_N), \qquad \sum_{j=1}^NS_j = K, \qquad \sum_{j=1}^NjS_j = N.
\end{align*}
Then, it specifies the distribution of $(s_1,\ldots,s_n)$ as the marginal distribution of $(S_1,\ldots,S_N)$ where the frequency counts of the last $N-n$ observations are marginalized out. If we in addition assume that $(S_1,\ldots,S_N)$ follows Ewens-Pitman's sampling formula as in \cite{hoshino2001applying} with parameters $\alpha\in[0,1)$ and $\theta>-\alpha$, which is equipped with sampling-consistency, the probability mass function is given as follows:
\begin{align}
    P(S_1,\ldots,S_N) = \frac{N!}{S_1!\cdots S_N!}\frac{\theta^{K\uparrow\alpha}}{\theta^{N\uparrow}}\prod_{j=1}^N\Bigg(\frac{(1-\alpha)^{j-1\uparrow}}{j!}\Bigg)^{S_j}\label{eq:PEwensPitman_Pop}
\end{align}
where $x^{k\uparrow a}$ for real numbers $x$ and $a$, and a nonnegative integer $k$ is defined as 
\begin{align}
    x^{k\uparrow a} = \begin{cases}
        1 &\text{for } k = 0\\
        x(x+a)\cdots(x+(k-1)a) & \text{for } k=1,2,\ldots,
    \end{cases}\nonumber
\end{align}
with $x^{k\uparrow}:= x^{k\uparrow 1}$. Because of the sampling consistency, the distribution of $(s_1,\ldots,s_n)$ has the same form as \eqref{eq:PEwensPitman_Pop} with $N,K$ and $(S_1,\ldots,S_N)$ replaced with their values in sample $n,k$ and $(s_1,\ldots,s_n)$. The parameter $\alpha$ is commonly called the discount parameter, which controls the extent of the power-law behavior in both tails of the distribution. With $\alpha$ closer to 1, the greater the power-law effect becomes, resulting in a partition having many parts with small partition sizes (i.e., proportions of $s_1, s_2$ and perhaps $s_3$ relative to $n$ increases) and one (or a few) giant part(s) with disproportionately high partition size(s). On the other hand, the parameter $\theta$, commonly called a strength parameter, controls the total number of parts in the partition, with greater $\theta$ resulting in more parts. 

The advantage of using Ewens-Pitman's partition structure (or any other sampling-consistent probability distributions on partitions of data) is that it assumes that observed cross-classifications are a subset of all possible cross-classifications that exist in the population. Of all possible combinations of elements across named entity categories, only a subset is likely present in the data. Therefore, a more conventional approach of fitting a model to the observed contingency table will fail to account for unseen named entities in a sample that may still exist in the population.
\subsubsection{Construction of the Risk Measure}
To construct a risk measure $DR(\mathscr{D}_\text{post}(\mathcal{R}))$, we need an estimate of the number of population uniques $\mathbb{E}(S_1)$. As $N$ is unknown but an extremely large number in our setting, an asymptotic version of $\mathbb{E}(S_1)$ will suffice. In \cite{hoshino2001applying}, $\mathbb{E}(S_1)$ was obtained using recursions of expectations of $K$ and its asymptotic version was derived through Stirling approximation. In this work, we derive this quantity directly. First, by considering the special case of $k_1 =1$ and all others to zero for the factorial moment of order $(k_1,k_2,\ldots,k_N)$ for frequency counts $(S_1,\ldots,S_N)$ of the Ewens-Pitman's sampling formula given in \citet[~Equation 4.11]{greve2025new}, we obtain the following:
\begin{align}
    \mathbb{E}(S_1) &= N\frac{(\theta+\alpha)^{N-1\uparrow}}{(\theta+1)^{N-1\uparrow}} \label{eq:EstPopUnique1}\\&= N\frac{\big[\frac{t^{N-1}}{(N-1)!}\big](1-t)^{-\theta-\alpha}}{\big[\frac{t^{N-1}}{(N-1)!}\big](1-t)^{-\theta-1}}\label{eq:EstPopUnique2}\\
    & =N\frac{[t^{N-1}](1-t)^{-\theta-\alpha}}{[t^{N-1}](1-t)^{-\theta-1}}\label{eq:EstPopUnique3}
\end{align}
where from \eqref{eq:EstPopUnique1} to \eqref{eq:EstPopUnique2}, we use the fact that $(N-1)$th coeffient of an exponential generating function $(1-t)^{-\theta-\alpha}$ is $(\theta+\alpha)^{N-1\uparrow}$ (note that the term $\big[\frac{t^{N-1}}{(N-1)!}\big]$ is an operator that takes the coefficient of $\frac{t^{N-1}}{(N-1)!}$ of a formal power series $f(t)$) and a similar result holds for the denominator $(\theta+1)^{N-1\uparrow}$. From \eqref{eq:EstPopUnique2} to \eqref{eq:EstPopUnique3}, we used a trivial fact that $N![t^N]f(t) = \big[\frac{t^N}{N!}\big]f(t)$ holds for any formal power series $f(t)$. Then, the value of $\mathbb{E}(S_1)$ given as \eqref{eq:EstPopUnique3} for large $N$ can be easily obtained using a well-known result in singularity analysis. In \citet[~Proposition 1]{flajolet1990singularity}, the asymptotic form of $[t^N](1-t)^{-\beta}$ for any $\beta\in\mathbb{R}$ with $\beta \neq0,-1,-2,\ldots$ is given as $\frac{N^{\beta-1}}{\Gamma(\beta)}$ (lower order terms for $N$ can also be easily obtained, but we only include $N^{\beta-1}$). Hence, we obtain the following
\begin{align}
    E(S_1) \approx \frac{\Gamma(\theta+1)}{\Gamma(\theta+\alpha)}N^\alpha\label{eq:AsympEstPopUnique}
\end{align}
for sufficiently large $N$. We use the right-hand side of \eqref{eq:AsympEstPopUnique} as an estimator of the number of population uniques $\hat{S}_1$. With this estimator $\hat{S}_1$, \cite{hoshino2001applying} proposes the proportion of population uniques in the sample relative to the number of sample uniques given by:
\begin{align}
    \hat{p} = \frac{\hat{S}_1}{s_1}\frac{n}{N}.\label{eq:PropPopUnqSmpUnq}
\end{align}
as one way to measure the risk. Note that this risk measure may be further improved by drawing references (e.g.,\citealp{favaro2016rediscovery,arbel2017bayesian} ) from species sampling literature that give estimates of the growth of the low frequency counts for additional $N-n$ samples through the estimation of discovery probabilities of low frequency counts.

The risk measure we propose utilizes the estimated proportion of population uniques relative to sample uniques $\hat{p}$ in \eqref{eq:PropPopUnqSmpUnq} to improve the following simple risk measure for swapping:
\begin{align*}
    1- \frac{\#\{\text{swapped $s_1$ in } \textbf{z}_J\}}{s_1}
\end{align*}
where the numerator is the number of swapped sample uniques $s_1$ in the pre-swap contingency table $\textbf{z}_J\in\mathscr{D}_\text{pre}$. Note that this is a function of post-swap contingency table $\tilde{\textbf{z}}_J\in\mathscr{D}_\text{post}(\mathcal{R})$ relative to the pre-swap contingency $\textbf{z}_J\in\mathscr{D}_\text{pre}$ and it is normalized to 1 if $r=0$ where we have $\mathscr{D}_\text{post}(\mathcal{R}) = \mathscr{D}_\text{pre}$. This simple risk measure evaluates the relative decrease in the number of sample uniques as a result of swapping. However, not all swapped sample uniques are population uniques that are the true subject of anonymization. For example, if we swap once and both of the swapped cells in $\textbf{z}_J$ happened to be sample uniques. Then, if for example $s_1 = 100$ for $\textbf{z}_J$, the (relative) risk becomes 0.98, with 2\% reduction in risk from the pre-swap data $\mathscr{D}_\text{pre}$. However, if the proportion of population uniques in the sample relative to sample uniques $\hat{p}$ is for instance 0.5, it is reasonable to assume that in expectation, only one of the two swapped sample uniques were a population unique which is subject to disclosure risk, resulting in 1\% reduction in risk. To account for this, we propose a risk measure given as follows:
\begin{align}
    DR(\mathscr{D}_\text{post}(\mathcal{R})) = 1- \Big(\frac{\#\{\text{swapped $s_1$ in } \textbf{z}_J\}}{s_1}\Big)\hat{p}.\label{eq:RiskMeasure}
\end{align}
Another way to look at this risk measure is that it is not only based on the reduction of the number of sample uniques $s_1$ in $\textbf{z}_J$, but also on the overall distribution on cell frequency counts $(s_1,\ldots,s_n)$ in $\textbf{z}_J$ which gives rise to the estimate of population uniques $\mathbb{E}(S_1)$ and thereby $\hat{p}$ through estimates of $\theta$ and $\alpha$. For practical implementation of the risk measure given as \eqref{eq:RiskMeasure}, we use the Newton-Raphson method to obtain MLE estimates $\hat{\theta}$ and $\hat{\alpha}$ needed to compute $\hat{p}$.
\subsection{Utility Measure}
For the utility measure, \cite{GomatamDecision2005} uses the difference in log likelihoods (i.e., logged version of the likelihood ratio) between post-swap and pre-swap contingency tables $\tilde{\textbf{z}}\in\mathscr{D}_\text{post}(\mathcal{R})$ and $\textbf{z}\in\mathscr{D}_\text{pre}$ evaluated with respect to the saturated model fitted to $\textbf{z}$. We instead use the likelihood of the mixture model fitted to the pre-swap embedding vector $\textbf{x}$ (which we used in Section \ref{subsec:ClustWeighting} for the weighting scheme) when constructing the data utility measure $DU$. In other words, we assume that the mixture model we fitted to pre-swap embedding vector $\textbf{x}$ approximates the true data-generating mechanism of the corpus of text chunks $\textbf{y}\in\mathscr{D}_\text{pre}$. Therefore, the difference in log likelihood between the post-swap embedding vector $\tilde{\textbf{x}}=(\tilde{x}_1,\ldots,\tilde{x}_n)$ and the pre-swap embedding vector $\textbf{x}$ evauated with respect to the mixture model on $\textbf{x}$ approximates the difference in log likelihood between post-swap text chunks $\tilde{\textbf{y}}\in\mathscr{D}_\text{post}(\mathcal{R})$ and the pre-swap text chunks $\textbf{y}\in\mathscr{D}_\text{pre}$ evaluated under the true data-generating mechanism. This difference can be considered a change in data utility.

Procedurally, we first compute the embedding vector $\tilde{\textbf{x}}$ from the post-swap text chunks $\tilde{\textbf{y}}$, which differ from the pre-swap embedding vector $\textbf{x}$ in $rn$ (note that this is always an integer) entries. Then, by evaluating the difference in the log likelihood between entries in $\tilde{\textbf{x}}$ that differ from $\textbf{x}$, which we denote $\tilde{\textbf{x}}\setminus\textbf{x}$ and the reverse  $\textbf{x}\setminus\tilde{\textbf{x}}$ under the fitted mixture model, we evaluate the data utility. More precisely, let
\begin{align*}
    \tilde{\textbf{x}}\setminus\textbf{x} = (\tilde{x}'_1,\ldots\tilde{x}'_{rn}), \qquad \textbf{x}\setminus\tilde{\textbf{x}} = (x'_1,\ldots,x'_{rn})
\end{align*}
and 
\begin{align*}
    \textbf{l'} &=(l'_1,\ldots,l'_{rn}), \qquad l'_j = \argmax_{k\in\{1,\ldots,K\}}\frac{\hat{\pi}_kf(x'_j|\hat{\phi}_k)}{\sum_{k=1}^K\hat{\pi}_kf(x'_j|\hat{\phi}_k)},\qquad j = 1,\ldots,rn\\
    \textbf{l} &=(l_1,\ldots,l_{n}), \qquad l_j = \argmax_{k\in\{1,\ldots,K\}}\frac{\hat{\pi}_kf(x_j|\hat{\phi}_k)}{\sum_{k=1}^K\hat{\pi}_kf(x_j|\hat{\phi}_k)},\qquad j = 1,\ldots,n
\end{align*}
be the vectors of cluster membership under the MLE estimates $(\hat{\pi}_1,\ldots,\hat{\pi}_K)$ and $(\hat{\phi}_1,\ldots,\hat{\phi}_K)$ of the mixture of PKB or spherical Cauchy distribitions for $rn$ observations that received swapping $\textbf{l}'$ and for all observations $\textbf{l}$. That is, $l'_j = k$ means that $\tilde{x}_j'$ and $x'_j$ belong to the $k$th cluster and same holds for $l_j$. Then, the difference in log likelihood between post-swap embedding vector $\tilde{\textbf{x}}$ and pre-swap embedding vector $\textbf{x}$ conditional on the cluster membership vector \textbf{l} is given as
\begin{align}
\log\big(f(\tilde{\textbf{x}}|\textbf{l},\hat{\phi}_1,\ldots,\hat{\phi}_K)\big) - \log\big(f(\textbf{x}|\textbf{l},\hat{\phi}_1,\ldots,\hat{\phi}_K)\big)
    =\sum_{i=1}^{rn} \log\big(f(\tilde{x}'_i|\hat{\phi}_{l'_i})\big)-\log\big(f(x'_i|\hat{\phi}_{l'_i})\big).\label{eq:DifLoglik}
\end{align}
where $f(\tilde{\textbf{x}}|\textbf{l},\hat{\phi}_1,\ldots,\hat{\phi}_K)$ and $f(\textbf{x}|\textbf{l},\hat{\phi}_1,\ldots,\hat{\phi}_K)$ are likelihoods of $\tilde{\textbf{x}}$ and $\textbf{x}$ under the mixture model used for weighting in Section \ref{subsec:ClustWeighting} conditional on the cluster membership $\textbf{l}$. Taking their log difference amounts to evaluating the difference in conditional (on $\textbf{l}'$) log-likelihoods for embeddings on observations that received swapping post and pre-swapping, resulting in the right hand side of \eqref{eq:DifLoglik}. 

To be in line with the range of data risk measure $DR$, we normalize the quantity \eqref{eq:DifLoglik} and use that as the data utility measure $DU$. Speicfically, we propose a data utility measure given as follows:
\begin{align}
    DU(\mathscr{D}_\text{post}(\mathcal{R})) &= 1+ \frac{\log\big(f(\tilde{\textbf{x}}|\textbf{l},\hat{\phi}_1,\ldots,\hat{\phi}_K)\big)-\log\big(f(\textbf{x}|\textbf{l},\hat{\phi}_1,\ldots,\hat{\phi}_K)\big)}{\log\big(f(\textbf{x}|\textbf{l},\hat{\phi}_1,\ldots,\hat{\phi}_K)\big)}\label{eq:UtilityMeasure1}\\
    &=\frac{\log\big(f(\tilde{\textbf{x}}|\textbf{l},\hat{\phi}_1,\ldots,\hat{\phi}_K)\big)}{\log\big(f(\textbf{x}|\textbf{l},\hat{\phi}_1,\ldots,\hat{\phi}_K)\big)}.\label{eq:UtilityMeasure2}
\end{align}
That is, it's a ratio between the log likelihood of the post-swap embedding vector $\tilde{\textbf{x}}\in\mathscr{D}_\text{post}(\mathcal{R})$ conditional on the estimated cluster membership \textbf{l} and that of the pre-swap embeddings $\textbf{x}\in\mathscr{D}_\text{pre}$. This measure is more numerically stable than simply taking the likelihood ratio between $\tilde{\textbf{x}}$ and $\textbf{x}$ by exponentiating Equation \eqref{eq:DifLoglik}. 

The embedding vectors  $ \tilde{\textbf{x}}\setminus\textbf{x}$ in post-swapped data $\mathscr{D}_\text{post}(\mathcal{R})$ is generated by chunks that contain swapped terms that distort the semantic meaning of the original chunks. Therefore, it (usually) has a lower likelihood than the corresponding embedding vectors $\textbf{x}\setminus\tilde{\textbf{x}}$ in pre-swapped data $\mathscr{D}_\text{pre}$ generated from original chunks. For this reason, the difference in the term \eqref{eq:DifLoglik} is negative, which makes the data utility measure less than one for $r> 0$. Note that depending on the choice of embedding models and the overall fit of the mixture model, the term \eqref{eq:DifLoglik} could take positive values, especially for small $r$. However, we rule out this possibility as an extreme case and assume $DU:\mathscr{D}_\text{post}(\mathcal{R})\to[0,1]$.

With the data risk measure defined as \eqref{eq:RiskMeasure} and the data utility as \eqref{eq:UtilityMeasure2}, the efficient frontier $\partial\mathscr{R}_\text{cand}$ for the decision problem covered in Section \ref{subsec:OptimizationProblem} will be on the lower right hand side of the figure if we put the utility measure $DU$ on x-axis and risk measure $DR$ on y-axis as in \cite{GomatamDecision2005}. We stick to their setting in the subsequent empirical example to determine the right optimal release $\hat{\mathcal{R}}$.
\section{Application: Prevention of Company Name Disclosure from Earnings Call Transcripts}\label{sec:Application}
We demonstrate the proposed methodology using a corpus of earnings call transcripts from the first quarter of the year 2018 in the information sector under NAICs (see Appendix \ref{appendix:Data} for details of data) with a total of $m=35$ earnings calls. The objective is to partially prevent the disclosure of company names through named entity swapping. Resulting anonymized text chunks could, for example, be used to carry out sentiment analysis on historical earnings call data without the LLM using a preconceived image of the company and/or a known scandal of the company for documents that predate such information. 

First, we provide a step-by-step introduction to the data preprocessing workflow that must take place to convert the corpus of earnings call texts into a contingency table of categorical variables $\textbf{z}$ and a vector of text chunks $\textbf{y}$. The latter is then transformed into a vector of embeddings $\textbf{x}$ via a text embedding model with dimension $d$. Throughout this application, we use OpenAI's text embedding model ``text-embedding-3-small'' with $d = 1536$. For the named entity recognition, we use the package \texttt{spaCy} \citep{spacy} with an English transformer pipeline model ``en\_core\_web\_trf''. The selection of embedding models and named entity recognition software is mostly based on availability and ease of implementation rather than any domain-specific knowledge about their suitability for our particular application. The main objective of this empirical analysis is to serve as a demonstration of the proposed named entity swapping procedure. As such, we do not attempt to change embedding models and named entity recognition software for a further increase in the anonymization performance. Finally, note that the majority of the data preprocessing steps covered are not specific to this application and can be readily applied to other forms of text data. 
\subsection{Data Preprocessing}\label{subsec:DataPreProcessing}
\subsubsection{Extraction of the Presentation Part from Earnings Call Transcripts}
Most earnings calls consist of two parts: a management presentation and a Q\&A session with analysts from banks and investment firms. The former is typically highly scripted and covers recurring topics such as profitability, cash flow, acquisitions, and the company’s or broader economy’s future outlook. These topics tend to be similar across companies. The latter, in contrast, is highly conversational. As a result, the semantic similarity between Q\&A sections of different earnings call transcripts is relatively low compared to the more structured presentation parts. Since our named entity swapping algorithm relies on semantic similarity between pairs of text chunks from different transcripts, the Q\&A portion is expected to contribute little to this process. Therefore, only the management presentation part of each transcript is used in the subsequent analysis.

\subsubsection{Semantic Chunking of Earnings Call Presentations}\label{subsec:chunking}

\begin{figure}
    \centering
    \includegraphics[width=0.9\linewidth]{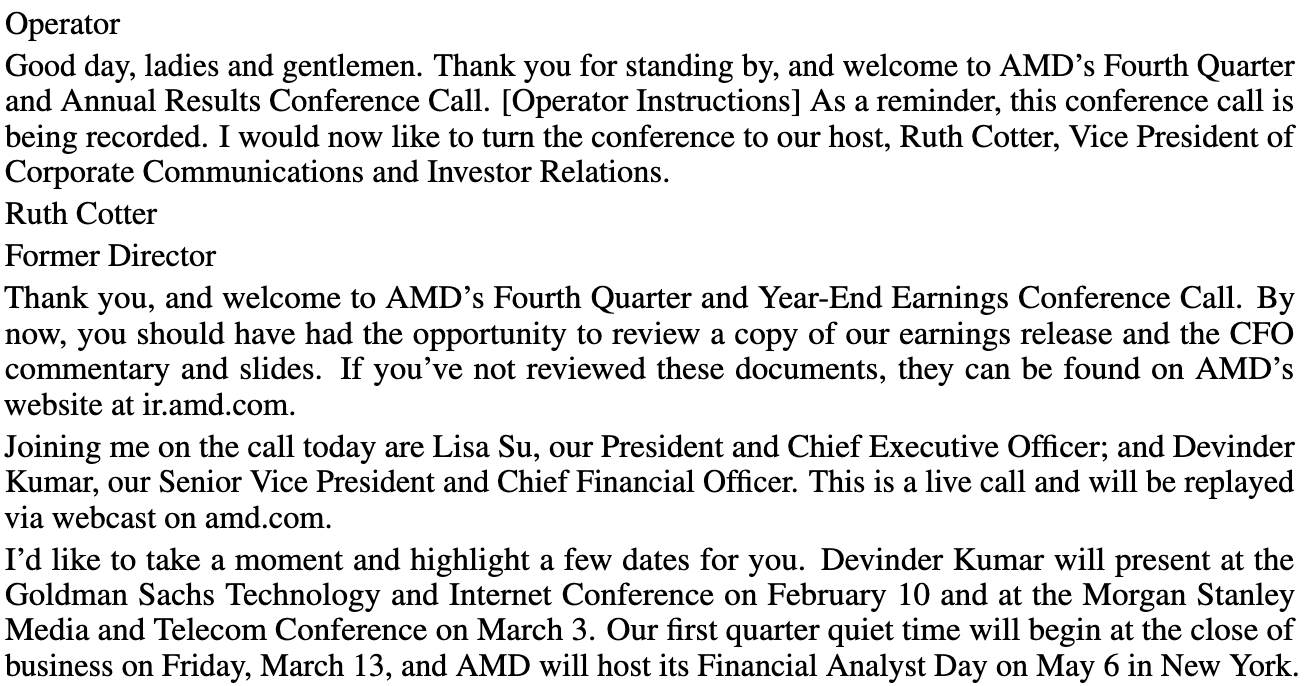}
     \caption{AMD Q4 2014 Earnings Call Opening}
     \label{fig:AMD_NoPreP}
\end{figure}

Once the presentation part is extracted, we proceed with partitioning it into nonoverlapping semantically uniform text chunks. Each chunk contains one or more consecutive sentences in the text that, when combined together, form a narrative that can be interpreted as one of the topics discussed in the text. The chunking is achieved through the use of a breakpoint-based semantic chunker \citep{llamaindex_semantic_chunking}. This approach uses the proximity between two consecutive sentences in the embedding space to determine whether a breakpoint should be inserted between them or not. Although this approach only uses the similarity between the two consecutive sentences and does not take into account a more gradual shift in the narrative, it results in a more semantically uniform chunk than a conventional fixed-size chunker that splits a document into fixed-sized chunks. For example, Figure \ref{fig:AMD_NoPreP} is an opening of the aforementioned (in Section \ref{subsec:NES}) Advanced Micro Devices earnings call presentation of the fourth fiscal quarter 2014.

The semantic splitter with the breakpoint threshold of 75 (it ranges between 0 to 100, with a higher threshold requiring greater semantic difference between the two sentences to split, resulting in bigger chunks) will insert two breakpoints between sentences in Figure \ref{fig:AMD_NoPreP}, resulting in three chunks.  The first chunk is the part ``Good day, \ldots Annual Results Conference Call'' while the second chunk is ``[Operator Instructions] As a reminder, \ldots via webcast on amd.com.'', and the final one starts with ``I'd like to take a moment'' till the end. Manual splitting will likely result in a slightly different split, but the overall semantic uniformity this procedure achieves within each chunk is acceptable. Note that in the above example, each paragraph may be considered a chunk, but not all parts of earnings call presentation transcripts have such well-separated paragraphs. Hence, this automatic chunking step is required. This procedure introduces one hyperparameter, which is the aforementioned breakpoint threshold. For our corpus of texts, it produced 1819 chunks in total. Hence, $n$ is set to 1819 in the subsequent analysis.
\subsubsection{Suppression of Direct Identifiers}\label{subsec:DI_Sup}
Direct identifiers are information that uniquely maps to the sensitive metadata we aim to protect from disclosure. All direct identifiers that reveal the company name should be suppressed prior to swapping. Direct identifiers in Figure \ref{fig:AMD_NoPreP} are URLs such as ir.amd.com and amd.com, as well as the company name and its abbreviation AMD. These are suppressed to ``[URL]'' and ``[Company Name]''. Executive names, such as Lisa Su, in principle do not meet the criteria for being a direct identifier as they have tenure. Therefore, without knowing the specific year in which the earnings call took place, one cannot uniquely identify the company name from the executive names. However, since we deal with the cross-section of earnings calls in a given quarter, the executive names will practically become a direct identifier. Hence, we suppress this information as well. Procedurally, this simply amounts to matching names in the chunk from the list of executives given in our data set and suppressing them as ``[Executive \#]'' where \# is the index for the given executive. 

The resulting chunks of text with all direct identifiers suppressed are shown in Figure \ref{fig:AMD_PostPreP}. It is this vector of text chunks with direct identifiers suppressed that is denoted as $\textbf{y} =(y_1,\ldots,y_n), n = 1819$. The accompanying metadata information we aim to anonymize through swapping is given as a vector of categorical variables $\textbf{c} = (c_1,\ldots,c_n)$ where each of the first $n_1$ entries $(c_1,\ldots,c_{n_1})$ equals the metadata category of the first document in the corpus with $n_1$ chunks and so on. For example, if the first document in the corpus is the aforementioned AMD earnings call, and since our goal is to anonymize the company name, we have $(c_1,\ldots,c_{n_1}) = (\text{AMD},\ldots,\text{AMD})$ and $y_2 = \text{``[Operator Instructions] As a reminder,\ldots via webcast on [URL]''}$.

From the vector of text chunks $\textbf{y} = (y_1,\ldots,y_n)$ with all direct identifiers removed, we use the text embedding model of our choice (which is OpenAI's ``text-emedding-3-smal'' with $d=1536$ in our application) to obtain the vector of text embeddings $\textbf{x} = (x_1,\ldots,x_n)$ where for all $i$ we have $x_i\in S^{d-1}, d = 1536$. For the numerical stability of the EM algorithm, it is advised that $d$ is much smaller than $n$, which is not the case in this example. However, this issue did not manifest in our subsequent demonstration of examples for the hyperparameter settings that were tried.
\subsubsection{Extraction of Named Entities for each Chunk}
Once a vector of text chunks (with direct identifiers removed) $\textbf{y}$ is given, one must convert it into a contingency table of named entity categories $\textbf{z}$ to enable the subsequent named entity swapping procedure covered in Section \ref{sec:DecThrNamedEntitySwapping}. To facilitate this, the user must provide all named entity categories that may potentially become quasi-identifiers. In this application, we consider $p=6$ named entity categories:
\begin{align}
    (\texttt{Organization}, \texttt{Person}, \texttt{Event},\texttt{Product},\texttt{Location},\texttt{Date}).\label{eq:NamedEntityCategs}
\end{align}
Our choice is mainly based on the categories supported by the package \texttt{spaCy} \citep{spacy} to carry out the named entity recognition. Recently, an LLM-based extension to \texttt{spaCy} called \texttt{spacy-llm} has also been introduced, which allows for more flexible categorization of named entities. As the aim of our work is mostly demonstrative, we do not consider using this extension, which is more flexible but requires careful tuning.

Once all named entity categories that may act as quasi-identifiers are specified, one can detect the presence of terms that belong to each category for each chunk. Note that a chunk may contain several terms that belong to the same category. For example, in the last chunk of Figure \ref{fig:AMD_PostPreP}, there are two terms ``Goldman Sachs'' and ``Morgan Stanley'' that belong to the category \texttt{Organization}. In that case, we treat the third chunk as having a value ``Goldman Sachs \& Morgan Stanley'' for a categorical variable \texttt{Organization}. This results in a sparse 6-way contingency table. It is possible to use another typical procedure in statistical disclosure control called recoding, where several categorical variables get merged into one class to resolve this problem. In this case, if there are other chunks with a term ``Goldman Sachs'' or  ``Morgan Stanley'', we can combine these variables into ``Investment Bank'' and also include the chunk with the variable ``Goldman Sachs \& Morgan Stanley'' into this category. However, we do not attempt this and keep the sparse contingency table as it is.

The resulting $6$-way contingency table of counts of text chunks is denoted as
\begin{align*}
    \textbf{z} = \{z_{i_1\ldots i_6}; i_j=1,\ldots,k_j; j = 1,\ldots,6\}
\end{align*}
where $z_{i_1\ldots i_6}$ represents the number of text chunks that consist of named entities
$
(i_1,\ldots,i_6) \in (\texttt{Organizaiton},\ldots,\texttt{Date}).
$ Hence, $\sum_{i_1,\ldots,i_6} z_{i_1 \ldots i_p} = n = 1819$. We combine the resulting contingency table $\textbf{z}$ obtained in this section with the vector of substrings $\textbf{y}$ and its embeddings $\textbf{x}$ in Section \ref{subsec:DI_Sup}, to define the pre-swap data
\begin{align*}
    \mathscr{D}_\text{pre} = \{\textbf{x},\textbf{y},\textbf{z}\}.
\end{align*}
This $\mathscr{D}_\text{pre}$ will receive named entity swapping in Section \ref{sec:DecThrNamedEntitySwapping} to produce post-swap data $\mathscr{D}_\text{post}(\mathcal{R})$ of which we pick the optimal $\mathscr{D}_\text{post}(\hat{\mathcal{R}})$ determined via solving the optimization problem in Section \ref{subsec:OptimizationProblem}.
\subsection{Empirical Results}\label{subsec:EmpRslt}
Here, we show an empirical demonstration of the proposed named entity swapping workflow. Out of $p=6$ named entity categories, we consider all but \texttt{Event} and \texttt{Date} for swapping. The category \texttt{Date} is set to $F$ (must be fixed) to maintain the cross-sectional structure of the earnings tall transcripts. The category \texttt{Event} is set to $C$ (must change) as out of 1819 chunks, only 36 chunks contain an entity categorized as such. These entities differ a lot in characteristics and thus swapping them around will result in relatively high distortion to the data. Hence, we suppress all entities in \texttt{Event} to a placeholder value ``[Event]'' which amounts to setting this category to $C$. The rationale is that the total number of these entities is too small to have any meaningful decrease in data utility $DU$ relative to the risk of keeping them unchanged.

\subsubsection{Effects of Hyperparameters in the Selection of Optimal Release}\label{subsec:HyperParamChoice}
We start with an empirical investigation of the extent to which the hyperparameter setting affects the choice of the optimal release $\hat{R}$. Hyperparameters for our swapping procedure are: (1) Breakpoint threshold for semantic chunks, (2) The total number of observations $N$ in the population contingency table $\textbf{Z}$, (3) The mixture component distribution $f$ (4) The number of mixture components $K$, (5) The lower bound for mixture proportions $\epsilon$. In addition, as mentioned previously, we fix the choice of embeddings to OpenAI embedding ``text-embedding-3-small'' with $d = 1536$.

After some experimentations, changes in the breakpoint threshold are deemed to have little to no effect on the choice of the optimal release $\hat{\mathcal{R}}$ as long as it is not set too low or too high, where these two extreme cases result in unreasonably high or low number of chunks respectively. For the remaining four hyperparameters, we construct the following grid of $3\times2^3=24$ hyperparameter settings:
\begin{align*}
    N\in\{1.0\times10^{10},1.0\times 10^{20},1.0\times 10^{30}\} \times f\in\{\text{PKB},\text{sCauchy}\} \times K\in\{10,30\} \times \epsilon \in\{1.0\times 10^{-3}, 1.0\times 10^{-10}\}.
\end{align*}
The choice of $N$ affects the values of the risk measure $DR$ while the other three parameters control the swapping weights (and thereby the actual pairs that received swapping) and the utility measure $DU$. 

For $N$, the value of $1.0\times10^{10}$ is chosen as a conservative lower bound. Since there are 6 named entity categories, even 100 unique entities within each category will result in $\textbf{Z}$ having $1.0\times 10^{12}$ cells. While it is reasonable to expect that the population contingency table $\textbf{Z}$ is sparsely populated as in $\textbf{z}$, this will still result in $N$ likely being in a similar order to the total number of cells in \textbf{Z}. Thus, $1.0\times10^{10}$ serves as a lower bound for $N$. By following a similar reasoning, $N=1.0\times10^{30}$ is picked as a conservative upper bound. $K$ and $\epsilon$ are chosen so as to generally allow for cluster sizes to differ greatly between clusters. The case with $K=30$ lies near the boundary where the EM algorithm has convergence issues for some settings. In fact, of all 24 combinations of $(N,f,K,\epsilon)$, six cases 
\begin{align*}
    &(1.0\times10^{30} ,\text{sCauchy},30,1.0\times 10^{-3}),(1.0\times10^{30} ,\text{sCauchy},30,1.0\times 10^{-10})),
    (1.0\times10^{20} ,\text{PKB},30,1.0\times 10^{-3}),\\
    &(1.0\times10^{20} ,\text{sCauchy},30,1.0\times 10^{-3}),
    (1.0\times10^{10} ,\text{PKB},30,1.0\times 10^{-10}),
    (1.0\times10^{10} ,\text{sCauchy},30,1.0\times 10^{-3})
\end{align*}
resulted in the EM-algorithm for the mixture model not achieving convergence to reasonable local maxima. Hence, for this data, $K=30$ represents a conservative upper bound for the number of clusters when looking for a sparse clustering solution with uneven cluster sizes by setting small $\epsilon$.

For each of these hyperparameter settings, the candidate release space $\mathscr{R}_\text{cand}$ consists of 180 releases $\mathcal{R}$. Firstly, out of four named entity categories considered for swapping, which are \texttt{Organization}, \texttt{Person}, \texttt{Product} and \texttt{Location}, we consider all combinations of two of them to be swapped simultaneously (and all others set to $U$), resulting in 6 combinations. The swapping we perform is done in a sequential manner. That is, after swapping one pair, out of the remaining pairs that constitute a valid swap, we swap one more, and so on. In total, we perform 30 consecutive swaps for each case and record a sequence of $DR$ and $DU$ with a maximum length of 30. As the total number of chunks in this dataset $n$ is 1819, this amounts to trying out the 30 different swap rates of 
$$r = \frac{2}{1819}, \frac{4}{1819},\ldots, \frac{60}{1819}$$ 
for each of the 6 combinations of two named entity categories out of four, resulting in a candidate release $\mathscr{R}_\text{cand}$ space of cardinality 180. Note that not all release $\mathcal{R}$ in $\mathscr{R}_\text{cand}$ produces a post-swap data $\mathscr{D}_\text{post}(\mathcal{R})$. This is because for some combinations of named categories, there are not enough valid swaps to achieve the swap rate $r$ above certain values.

\begin{figure*}
    \centering
    \begin{subfigure}[t]{\textwidth}
        \includegraphics[width=\linewidth]{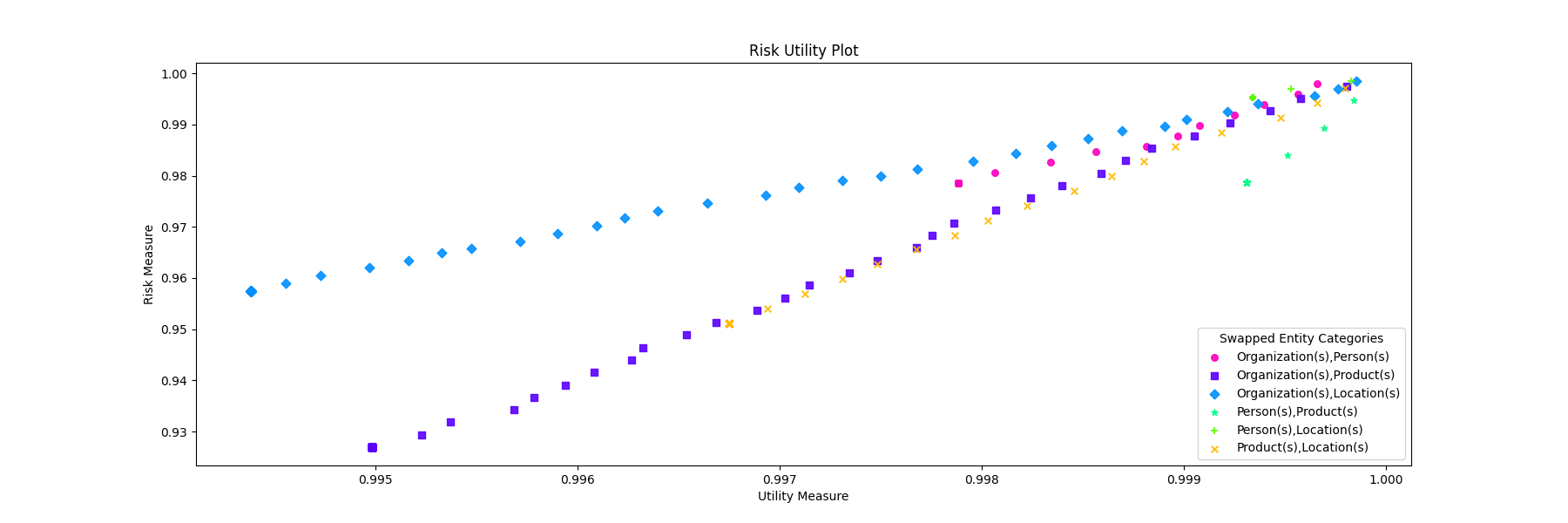}
        \caption{$N=1.0\times 10^{10}$, $f$=PKB, $K=10$ and $\epsilon = 1.0\times 10^{-3}$}
    \end{subfigure}
    \hfill
    \begin{subfigure}[t]{\textwidth}
        \includegraphics[width=\linewidth]{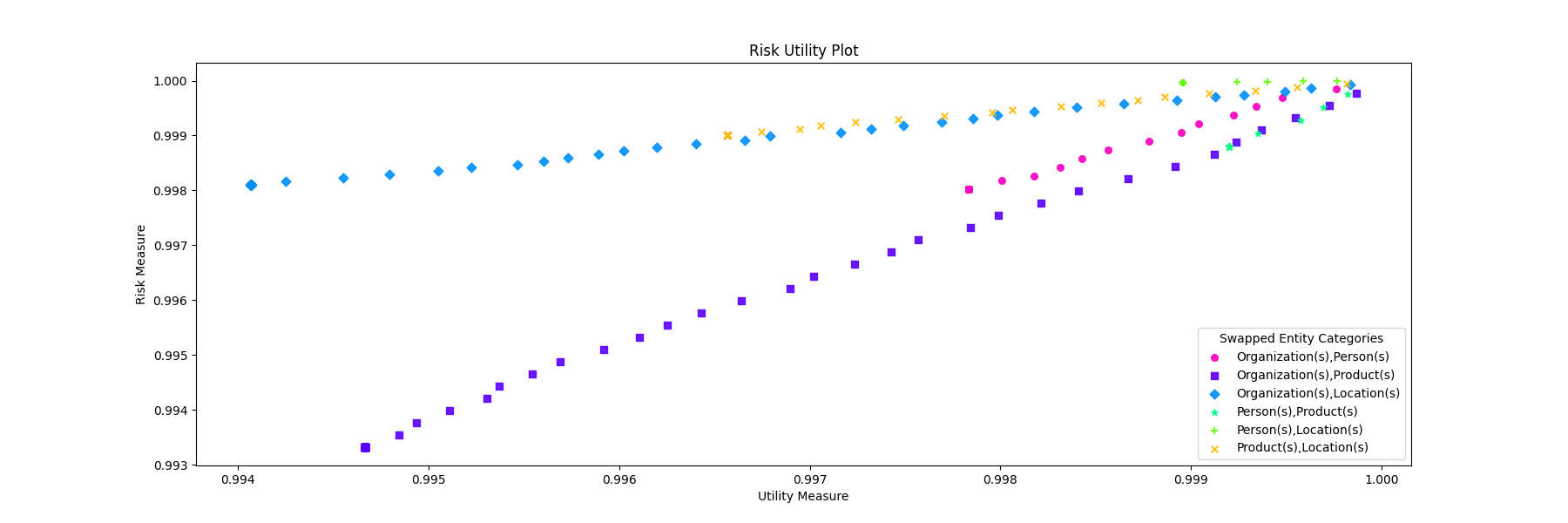}
        \caption{$N=1.0\times 10^{20}$, $f$=PKB, $K=10$ and $\epsilon = 1.0\times 10^{-3}$}
    \end{subfigure}
    \vspace{0.5cm}

    \begin{subfigure}[t]{\textwidth}
        \includegraphics[width=\linewidth]{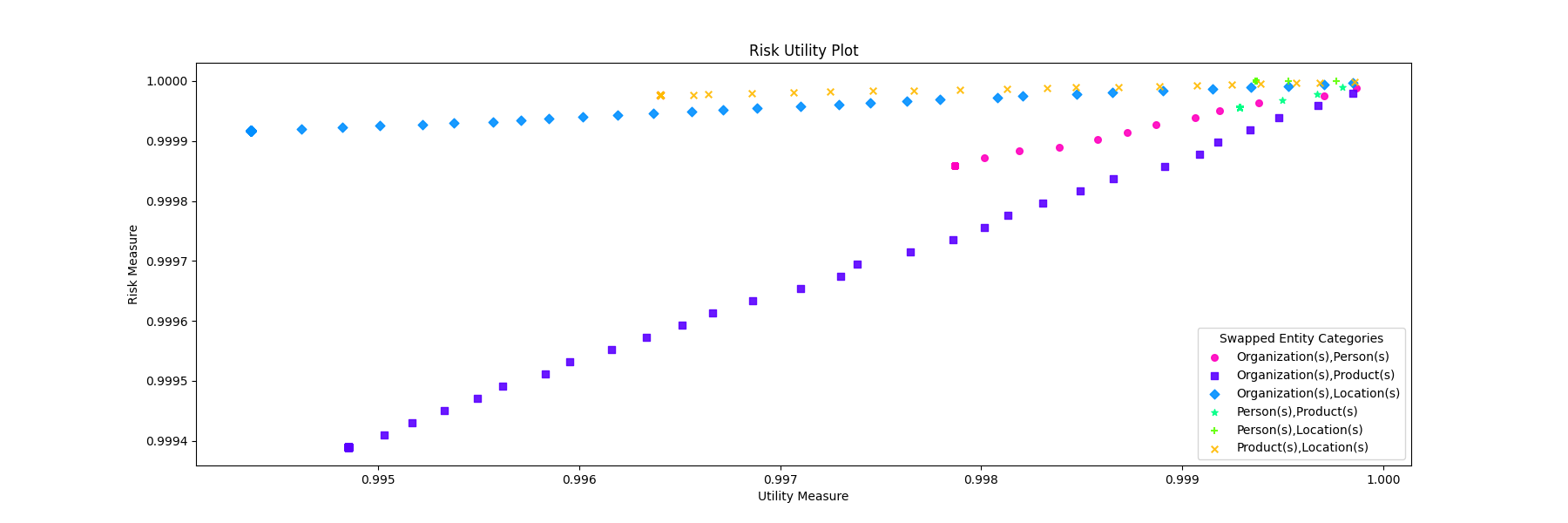}
        \caption{$N=1.0\times 10^{30}$, $f$=PKB, $K=10$ and $\epsilon = 1.0\times 10^{-3}$}
    \end{subfigure}
    \hfill
    \caption{Risk-Utility Plots for PKB Mixtures with $K=10, \epsilon = 1.0\times 10^{-3}$ and Three Different $N$}
    \label{fig:PKBwDifN}
\end{figure*}

Firstly, for a fixed $(f,K,\epsilon)$, we compare the effect that the population size $N$ has on picking the optimal release $\hat{\mathcal{R}}$. Figure \ref{fig:PKBwDifN} shows the risk utility plots when $f = \text{PKB}, K = 10$ and $\epsilon = 1.0\times10^{-3}$, and three different values of $N$ are used. Each figure shows $DR$ and $DU$ for all releases we consider using the aforementioned setting. That is, points at the top right around $(1,1)$ are when $r$ is set to $2/1819$, and for subsequent swap rates, points generally move towards the bottom left at different rates depending on the combination of named entity categories. 

In addition, Table \ref{tab:MLEthetaalpha} shows MLE estimates of parameters $\theta$ and $\alpha$ for the Ewens-Pitman sampling formula and the proportion of sample uniques for all six marginalized contingency tables: $\textbf{z}_{\texttt{Organization},\texttt{Person}},\ldots,\textbf{z}_{\texttt{Product},\texttt{Location}}.$

\begin{table}[ht]
\centering
\renewcommand{\arraystretch}{1.5} 
\begin{tabular}{lcccccc}
\toprule
 & \begin{tabular}
 {@{}c@{}}\texttt{Organization}\\\texttt{Person}\end{tabular} & 
   \begin{tabular}{@{}c@{}}\texttt{Organization}\\\texttt{Product}\end{tabular} & 
   \begin{tabular}{@{}c@{}}\texttt{Organization}\\\texttt{Location}\end{tabular} & 
   \begin{tabular}{@{}c@{}}\texttt{Person}\\\texttt{Product}\end{tabular} & 
   \begin{tabular}{@{}c@{}}\texttt{Person}\\\texttt{Location}\end{tabular} & 
   \begin{tabular}{@{}c@{}}\texttt{Product}\\\texttt{Location}\end{tabular} \\
   \midrule
$\hat{\theta}$ & 15.774 & 35.259 & 58.882 & 31.770 & 15.443 & 27.602 \\
$\hat{\alpha}$ & .889 & .896 & .864 & .865 &.769 & .831 \\
$s_1/n$ & .238 & .245 & .237 & .122 & .081 & .117\\
\bottomrule
\end{tabular}
\caption{Sample Unique ratio $s_1/n$ and MLE estimates of $\theta$ and $\alpha$.}
\label{tab:MLEthetaalpha}
\end{table}

Most strikingly, from Figure \ref{fig:PKBwDifN}, one can observe that when $N = 1.0\times 10^{10}$, the combination $\texttt{Product} =S, \texttt{Location} = S$  is part of the efficient boundary $\partial \mathscr{R}_\text{cand}$ together with the combination $\texttt{Organization} = S, \texttt{Product} = S$ for approximately the first 10 swaps (if we ignore the short-lived combination of $\texttt{Person} = S, \texttt{Product}=S$ which only reaches up to 5 swaps), and even reaching superior risk and utility tradeoff than the latter for the remaining swaps until it run out of swappable pairs reaching the utility of about 0.997 with the corresponding risk of about 0.95. This combination, however, for cases $N= 1.0\times 10^{20}$ and $N=1.0\times 10^{30}$ has the worst risk-utility trade-off among all six combinations that are tried. This is because the data risk measure $DR$ in \eqref{eq:RiskMeasure} balances the relative reduction in the number of sample uniques captured by the term $-\frac{\#\{\text{swapped $s_1$ in } \textbf{z}_J\}}{s_1}$ with the proportion of population uniques relative to the sample uniques captured by the term $\hat{p}$. The latter term, in addition to $N$, is parameterized by $\hat{\theta},\hat{\alpha}$ which are estimated using not only sample uniques $s_1$ but also other frequency counts $s_2,\ldots,s_n$. In case of the combination $\texttt{Product}=S, \texttt{Location} = S$, the ratio of sample unique to $n$ is one of the lowest, resulting in a greater magnitude of decrease in the term $-\frac{\#\{\text{swapped $s_1$ in } \textbf{z}_J\}}{s_1}, J = (\texttt{Product},\texttt{Location})$ compared to other combinations. However, as the estimated discount parameter $\alpha$, which controls the strength of the power-law behavior of Ewens-Pitman's sampling formula is smaller than most other combinations, the estimated proportion of population uniques relative to sample uniques $\hat{p}$ is low, which thereby reduces the magnitude of the former term. The scale of such reduction by $\hat{p}$ is controlled by $N$ and $\hat{\alpha}$ through the term $N^{\alpha-1}$ in \eqref{eq:PropPopUnqSmpUnq}. Thus, as $N$ gets larger, and since the estimated discount parameter $\hat{\alpha}$ is smaller than most other combinations, the greater the reduction in effect size becomes. Therefore, this combination performs the worst in cases $N= 1.0\times 10^{20}$ and $N=1.0\times 10^{30}$ despite performing well when $N=1.0\times 10^{10}$. 

Conversely, the combination $\texttt{Organization} = S, \texttt{Product}=S$ has the greatest risk-utility tradeoff in all three settings of $N$ despite having the greatest number of sample uniques $s_1$ which results in the term $-\frac{\#\{\text{swapped $s_1$ in } \textbf{z}_J\}}{s_1}, J = (\texttt{Organization},\texttt{Product})$ achieving smallest magnitude decrease. This is because the estimates $\hat{\theta}$ and $\hat{\alpha}$, which affect the term $\hat{p}$ have favorable values compared to all other combinations. This is firstly because the estimated discount parameter $\hat{\alpha}$ is closest to 1, which makes this distribution have the most power-law like behavior, resulting in a greater expected number of population uniques. In addition, compared to another combination $\texttt{Organization} = S, \texttt{Person} = S$ which also has a similar value for the discount parameter $\hat{\alpha}$, the combination $\texttt{Organization} = S, \texttt{Product}=S$ has a much greater estimate of the strength parameter $\hat{\theta}$ which also contributes to the increase in the expected number of population uniques. For these reasons, among all six combinations, $\texttt{Organization} = S, \texttt{Product}=S$ achives the greatest $\hat{p}$ which results in overcoming the fact that it has the smallest magnitude decrease in the term $-\frac{\#\{\text{swapped $s_1$ in } \textbf{z}_J\}}{s_1}$.

Now as $N = 1.0\times 10^{10}$ was a conservative lower bound for the unknown $N$, and the consistent performance of $\texttt{Organization} = S, \texttt{Product}=S$ mostly achieving the greatest risk-utility tradeoff amoung all combinations regardless of values of $N$, we focus only on cases $N=1.0\times 10^{20}$ and $N=1.0\times10^{30}$ and compare the effects of other hyperprameters $(f,K,\epsilon)$. Hence, out of 15 remaining hyperparameter settings (24 minus 6 nonconvergent cases and 3 shown in Figure \ref{fig:PKBwDifN}), we do not consider the 5 settings involving the case $N=1.0\times 10^{10}$. For all other settings, results are in Appendix~\ref{appendix:Figures4and5} as Figures~\ref{fig:PKBrest} and \ref{fig:sCrest}.

In Figure \ref{fig:PKBrest}, results of five of the remaining settings involving the use of PKB distribution as the component density $f$ are shown.  All these settings will yield the same conclusion that $\texttt{Organization} =S,\texttt{Product} = S$ will result in the best risk-utility tradeoff. Similarly, in Figure \ref{fig:sCrest}, five remaining settings involving spherical Cauchy mixtures are shown. Again, the conclusion is the same as in Figure \ref{fig:PKBrest}.

All in all, when $N=1.0\times 10^{20}$ or $N=1.0\times 10^{30}$, regardless of the other hyperparameter settings, the combination $\texttt{Organization} = S, \texttt{Product} = S$ results in the best trade-off between risk and utility for almost all values of $r$ we considered except when r is really small, then $\texttt{Person} = S, \texttt{Product} = S$ also has a comparable or even better trade-off for some hyperparameter settings. For this reason, the efficient boundary $\partial\mathscr{R}_\text {cand}$ consists almost exclusively of releases with $\texttt{Organization} = S, \texttt{Product} = S$. Therefore, regardless of the choice of the risk-utility tradeoff $a$ and offset $c$ in the risk-utility equation \eqref{eq:RU_LinEq}, the optimization problem given in Section \ref{subsec:OptimizationProblem} picks a release with swapped named entity categories set to  $\texttt{Organization}$ and $\texttt{Product}$ out of the candidate releases $\mathscr{R}_\text{cand}$.

\subsubsection{Predictictive Checks on Pre- and Post-swap Chunks}\label{subseq:PredictiveChecks}
From the results in Section \ref{subsec:HyperParamChoice}, it was determined that the selection of two named entity categories most suitable for swapping appears to be robust to the choice of clustering-related hyperparameters $f, K$ and $\epsilon$. On the other hand, it was also demonstrated that the choice of $N$ does play a major role in picking the best combination of two named entities. Based on the reasoning in Section \ref{subsec:HyperParamChoice}, $N=1.0\times 10^{10}$ is not a realistic choice of $N$ compared to other options. Hence, we determine that categories $\texttt{Organization}$ and $\texttt{Product}$, which are deemed most suitable when fixing $N=1.0\times 10^{20}$ or $N=1.0\times 10^{30}$ to be ideal candidates for swapping.

To pick the optimal release $\hat{\mathcal{R}}$, one must also specify the swap rate $r$. Unless the trade-off coefficient $a$ is picked so as to make the linear risk-utility equation coincide with the line of the risk-utility coordinates of releases with $\texttt{Organization} = S, \texttt{Product}=S$ with various values for $r$, the greatest possible $r$ will be the optimal choice for most of the reasonable choices of $a$ and $c$. That is, unless one has a large value of $a$ and an extremely negative offset $c$ to pick intermediate values of $r$ (this corresponds to having a strict utility requirement, the extreme case picking $r=0$ for sufficiently large $a$ and small $c$) or an almost flat $a$ with an offset $c$ close to 1 to again pick intermediate values of $r$ (this corresponds to having a strict risk requirement, the extreme case picking $r=0$ for $a\approx 0$ and $c\approx 1$), maximum achievable value of $r$ will be the optimal choice. The maximum possible value of $r$ is directly proportional to the total number of swaps that take place until all the valid pairs are exhausted. This is about 30-40 swaps for our data based on the hyperparameter choice $f,K$ and $\epsilon$ as well as the order in which the pairs are chosen for the consecutive swapping. Based on these reasonings, the resulting optimal release $\hat{\mathcal{R}}$ is given as:
$$
\hat{\mathcal{R}} = (\hat{r}=\max(r), \texttt{Organization} =S,\texttt{Person}=U,\texttt{Event}=C,\texttt{Product}=S,\texttt{Location}=U,\texttt{Date}=F).
$$

From this optimal release $\hat{\mathcal{R}}$, we randomly generate the post-swap data $\mathscr{D}^{(i)}_\text{post}(\hat{\mathcal{R}}) = \{\hat{\textbf{x}},\hat{\textbf{y}},\hat{\textbf{z}}\}, i = 1,\ldots,M$ to carry out the predictive checks on the anonymization performance. Specifically, for each draw $i\in\{1,\ldots,M\}$ we compare the post-swap chunks $\hat{\textbf{y}}^{(i)}\in\mathscr{D}^{(i)}_\text{post}(\hat{\mathcal{R}})$ with the pre-swap chunks $\textbf{y}\in\mathscr{D}_\text{pre}$ with respect to the predictability of their metadata vector $\textbf{c}$. We denote elements in post-swap chunks $\hat{\textbf{y}}^{(i)}$ that received swapping and thereby differ with pre-swap chunks $\textbf{y}$ as 
$
\hat{\textbf{y}}^{(i)}\setminus\textbf{y} 
$
and the corresponding elements in pre-swap chunks $\textbf{y}$ as 
$
\textbf{y}\setminus\hat{\textbf{y}}^{(i)}.
$
Both are vectors of text chunks with $rn$ elements where each chunk differs only in named entities categorized as $\texttt{Organization}$ and $\texttt{Product}$ between them. Both $\hat{\textbf{y}}^{(i)}\setminus\textbf{y}$ and $\textbf{y}\setminus\hat{\textbf{y}}^{(i)}$ are separately supplied to two state-of-the-art LLMs with an instruction (the specific prompt is given in the Appendix \ref{appendix:LLM_Prompt}) to predict the company name of this substring obtained from an earnings call transcript. The two LLMs we use for this task are Google Gemini 2.5 Flash and OpenAI GPT-4o, both with temperature (the randomness/creativity in response) set to 0. For other LLM specifications, refer to Appendix \ref{appendix:LLM_Prompt}. The corresponding entries in the metadata vector $\textbf{c}$ that represent the true company name for these chunks $\hat{\textbf{y}}^{(i)}\setminus\textbf{y}$ and $\textbf{y}\setminus\hat{\textbf{y}}^{(i)}$ are denoted as $\textbf{c}^{(i)}$. For the subsequent two experiments, we set $M = 30$.

The first experiment we perform is the comparison of the predictive accuracy of company names between pre-swap chunks $\textbf{y}\setminus\hat{\textbf{y}}^{(i)}$ and the post-swap chunks $\hat{\textbf{y}}^{(i)}\setminus\textbf{y}$ for $i = 1,\ldots M$, using two LLMs. Specifically, for each run $i$, we generate predictions $\textbf{c}^{(i)}_\text{pre,Gemini2.5}$ and  $\textbf{c}^{(i)}_\text{pre,GPT-4o}$ for $\textbf{c}^{(i)}$ using pre-swap chunks $\textbf{y}\setminus\hat{\textbf{y}}^{(i)}$ with Gemini 2.5 Flash and GPT-4o and compute the accuracy
\begin{align*}
    \frac{\#\{\textbf{c}^{(i)}_{\text{pre,llm}}=\textbf{c}^{(i)}\}}{rn}
\end{align*}
for $\text{llm} \in\{\text{Gemini2.5, GPT-4o}\}$ where $\#$ counts the number of true occurances. Similarly, the accuracy for the two predictions $\textbf{c}^{(i)}_\text{post,Gemini2.5}$ and  $\textbf{c}^{(i)}_\text{post,GPT-4o}$ for $\textbf{c}^{(i)}$ using post-swap chunks  $\hat{\textbf{y}}^{(i)}\setminus\textbf{y}$ are also computed. We compare the averages and standard deviations of these accuracies between pre-swap and post-swap chunks for each LLM by repeating this experiment $M=30$ times.

\begin{table}[h]
\centering
\begin{tabular}{lcc}
\toprule
 & \textbf{Pre-swap} & \textbf{Post-swap} \\
\midrule
\textbf{Gemini 2.5 Flash} & 0.934 (0.022) & 0.602 (0.042) \\
\textbf{GPT-4o} & 0.888 (0.018) & 0.564 (0.051) \\
\bottomrule
\end{tabular}
\caption{Pre- and post-swap averages and standard deviations (in brackets) of the prediction accuracy of the company names for two LLMs simulated with $M=30$}
\label{tab:AccuraciesPrePost}
\end{table}

Table \ref{tab:AccuraciesPrePost} shows the average accuracy and the corresponding standard deviation (in brackets) for $M=30$ runs of this experiment with $(f,K,\epsilon)$ set to $(\text{sCauchy},10,1.0\times10^{-10})$. Note that for optimal release $\hat{R}$ the source of randomness in mapping from $\hat{\mathcal{R}}$ to $\mathscr{D}^{(i)}_\text{post}(\hat{\mathcal{R}}), i = 1,\ldots, M$, is in the random sampling of swapping pairs conditional on the clustering weights. Hence, it depends on $(f,K,\epsilon)$ but not on $N$. For both LLMs, there is a significant decrease in the accuracy of the predictability of the company names after swapping. Furthermore, the variability in the accuracy is also small partly because the randomness in mapping from $\hat{\mathcal{R}}$ to $\mathscr{D}^{(i)}_\text{post}(\hat{\mathcal{R}}), i = 1,\ldots, M$ is small. This corroborates the earlier claim that the introduction of a clustering-based weighting scheme has contributed to reducing the randomness caused by this part. On average, $36.37$ pairs are swapped, resulting in the swap rate of approximately $4\%$.

The second predictive check we perform uses the aggregate pre- and post-swap predictions $\textbf{c}^{(1:30)}_\text{pre,Gemini2.5},\textbf{c}^{(1:30)}_\text{post,Gemini2.5},$ and $\textbf{c}^{(1:30)}_\text{pre,GPT4o},\textbf{c}^{(1:30)}_\text{post,GPT4o}$. The pre-swap and post-swap aggregate predictions were performed in a total of 2182 chunks for each LLM. The results are then arranged in $2\times2$ contingency tables as follows:

\begin{table}[h]
\centering
\hspace{-2.0cm}
\begin{subtable}[t]{0.4\textwidth}
\centering
\begin{tabular}{lcc}
\toprule
 & \textbf{Post-swap Correct} & \textbf{Post-swap Wrong} \\
\midrule
\textbf{Pre-swap Correct} & 1277 & 760 \\
\textbf{Pre-swap Wrong}   & 37  & 108 \\
\bottomrule
\end{tabular}
\caption{Gemini 2.5 Flash}
\end{subtable}%
\hspace{0.16\textwidth}%
\raisebox{-19pt}{\rule{1.0pt}{44pt}}%
\hspace{-0.01\textwidth}%
\begin{subtable}[t]{0.3\textwidth}
\centering
\begin{tabular}{lcc}
\toprule
 & \textbf{Post-swap Correct} & \textbf{Post-swap Wrong} \\
\midrule
& 1200 & 738 \\
& 31  & 213 \\
\bottomrule
\end{tabular}
\caption{GPT-4o}
\end{subtable}
\caption{Contingency tables comparing aggregate pre-swap and post-swap prediction results for each LLM}
\label{tab:2times2tables}
\end{table}

For both of the contingency tables in Table \ref{tab:2times2tables}, McNemar's test rejects the null hypothesis that pre-swap and post-swap binary classifiers are equal in any reasonable choice of significant levels.

Based on these two predictive checks, we conclude that the swapping has indeed significantly reduced the predictability of the company name for chunks that received swapping.

\subsubsection{Characteristics of Swapped Chunks}

As a final check, we take a detailed look at some of the swapped chunks to assess the overall characteristics of the named entity swapping. Firstly, we start with the case where both chunks were originally correctly identified, but after the swap, resulted in both being mislabeled by the LLMs, thus performing the intended function. These two post-swap chunks below are from a cluster of chunks that mostly consist of texts that describe sales of their products and services to another company:

\begin{quote}
    [URL] 5 The trend of system houses building their own silicon continues, with the chip design group at premier hyperscale web service provider adopting our software and hardware solutions for 7-nanometer designs. As we reported in Q3, we are collaborating with iC-Haus, ARM and TSMC to build the industry first test chip for IC Validator or CCIX, incorporating [Company Name] IP and using [Company Name] tools on the TSMC 7-nanometer FinFET process. 2017 was a year of rapid advancements for Automotive.
\end{quote}

\begin{quote}
     Success with advanced process technology also continues. One example was a competitive win on an ARM CPU implementation in China, driven by better total power consumption results. Reflecting the compelling benefits of integration of physical verification into our core digital flow, Xilinx, known for its ASICs industrial, automotive and medical technology, selected Cache Coherent Interconnect for Accelerators, replacing incumbent tools. Our custom analog product group also posted a strong quarter.
\end{quote}

The first chunk is from the earnings call of Cadence Design Systems, while the second one is from its market rival, Synopsys. Here, the swapping resulted in both LLMs confusing the former with the latter and vice versa. While this result may simply imply that pseudo-identifiers got swapped between these two chunks and LLMs have simply interchanged their predictions accordingly, the change in the structure of the information these two chunks experienced is most likely more complicated. The first pre-swap chunk from Cadence's earnings call consisted of the following named entities:
$$
\{\texttt{Organization} = \text{ARM \& Automotive \& TSMC \& Xilinx}, \texttt{Product}= \text{Cache Coherent Interconnect for Accelerators}\}.
$$
On the other hand, the second pre-swap chunk from Synopsis' earnings call consisted of the following named entities:
$$
\{\texttt{Organization}= \text{ic-Haus},\texttt{Location}= \text{China}, \texttt{Product}=\text{IC Validator}\}.
$$
A quick internet search will reveal that ``ARM'', ``TSMC'', ``Xilinx'' and ``Cache Coherent Interconnect for Accelerators'' (CCIX) will point to several announcements from these firms regarding their collaboration with Cadence. Hence, these terms act as quasi-identifiers of Cadence. Similarly, for the Synopsis chunk, ``IC Validator'' and ``ic-Haus'' act as quasi-identifiers. As the swapping with respect to terms in \texttt{Organization} and \texttt{Product} resulted in ``Xilinx'' and ``ic-Haus'' in \texttt{Organization} switching positions as well as ``Cache Coherent Interconnect for Accelerators'' and ``IC Validator'' in \texttt{Product} being interchanged, the first Cadence chunk now contains Synopsis' quasi-identifiers and part of Cadence's quasi-identifiers (in fact, the named entity recognition did not recognize the term CCIX which is an abbreviation of ``Cache Coherent Interconnect for Accelerators'', so this term remains in Cadence's chunk) while the second Synopsis chunk contains the other part of Cadence's quasi-identifiers. Therefore, the Synopsis chunk now only contains a relatively weak link to Cadence through terms ``Xilinx'' and `Cache Coherent Interconnect for Accelerators'', the latter being an open industry standard rather than Cadence's product, hence requiring other company names such as ARM, TSMC and in addition to Xilinx to meaningfully serve as quasi-idenfiers. For this reason, even though both LLMs did guess the second Synopsis chunk as belonging to Cadence's earnings call, most likely based on these entities sourced from Cadence's first chunk, the association is likely not so strong. In contrast, the first chunk belonging to Cadence's earnings call now contains two conflicting quasi-identifiers. Particularly ``IC Validator'' is in fact Synopsis' product, so this may have been instrumental in both LLMs determining that the first Cadence chunk belongs to Synopsis' earnings call. To summarize, after named entity swapping, two chunks with strong ties to the corresponding true metadata got turned into one chunk with strong but conflicting ties to two metadata and another chunk with a weak connection to the misleading metadata. Readers are encouraged to supply these post-swap chunks to their favorite choice of LLM and gradually reveal the extent of data suppression to determine the uncertainty in the LLM's predictions of these company names. For example, one may start with asking the first and second guesses for the company name, followed by revealing that some terms in these texts are altered, then finally revealing that in fact some terms are interchanged between these two chunks and record how the resulting guesses and certainty of these guesses evolve.

In comparison, the two post-swap chunks below are from a cluster of chunks that mostly deal with positive news about the company regarding revenue, operating income, etc. For these chunks, LLMs classified them correctly both pre- and post-swapping, thus anonymization did not succeed for these chunks:
\begin{quote}
    Moving on to King. Global Business King generated \$4.2 billion of revenue this quarter, up 1\% at actual rates and down 1.5\% at constant currency. We had modest growth in GBS signings, marking the fourth consecutive quarter of signings growth. And our GBS revenue was up in several regions, including Asia Pacific and Latin America. We have good momentum in Consulting but continue to see declines in Application Management, particularly in North America and Europe. Consulting revenue grew 1\% again this quarter. We've said for some time that the path to revenue growth starts with signings, which then translates to backlog growth. Our Consulting backlog was essentially flat in the second quarter and was up starting in the third. And we've now driven 2 consecutive quarters of Consulting revenue growth. Revenue in our Candy Crush and iX business grew about 40\%. And we're also seeing good growth in the new practices we've built around our innovative technologies like AI and Blockchain. 
\end{quote}

\begin{quote}
    'Each part of our business reached new milestones and demonstrated the durable and enduring nature of our franchises. [Company Name] celebrated strong Call of Duty momentum in their best operating income year ever. [Company Name] delivered their highest operating income ever for a year with no major game releases, and Services returned to growth with the Digital Strategy franchise stronger than ever. 2017 was also an important year for us for preparation of our future growth initiatives. 
\end{quote}

The first chunk belongs to IBM and the second to Activision Blizzard. A combination of several shortcomings in the existing implementation of the named entity swapping algorithm has made anonymization in these chunks insufficient. To start with, these chunks are from a cluster with relatively low concentration (= more variability in chunks), with chunks belonging to the same cluster covering wide topics such as revenue growth, product sales and other positive news about the company in general. Therefore, even though these chunks do not seem to have much in common except the overall narrative, they were matched as a pair. To resolve this issue, one may set a lower threshold for the concentration parameter of the cluster to prevent extracting any pairs from clusters with sufficiently low semantic similarity among chunks that belong to those. Another solution is to incorporate a more coarse industry classification as covariates in the clustering algorithm. In this way, one can prevent the pairing of firms that may share the same sector but serve different industries. In addition, especially for the first chunk, the shortcomings of the existing named entity recognition library \texttt{spaCy} we utilize are clear. Named entities in the first chunk from IBM are:
\begin{align*}
    \{&\texttt{Organization} = \text{Consulting \& GBS \& Global Business Services \& Services}, \\ &\texttt{Location}=\text{Asia Pacific \& Europe \& Latin America \& North America},\texttt{Product} = \text{Digital Strategy}\}
\end{align*}
For the second chunk from Activision Blizzard, those are:
\begin{align*}
    \{\texttt{Organization} = \text{King}, \texttt{Product}=\text{Call of Duty \& Candy Crush}\}
\end{align*}
Firstly, the named entity recognition missed an important term ``iX'', which is the name of the IBM digital consulting division and thus may act as a quasi-identifier combined with terms such as ``Consulting''. Secondly, terms such as ``AI'' and ``Blockchain'', even if those may not be considered a part of  \texttt{Product} should be categorized into something, as these could also act as quasi-identifiers (perhaps not these days as every company likes to talk about it). Hence, it also shows the limitation in the selection of named entity categories. These issues may be resolved using a more flexible LLM-based named entity recognition from \texttt{spacy-llm} library. The last problem relates to the current implementation of the swapping algorithm. Namely, when the number of entities in a named entity category that receives swapping differs between the pair, there will be some entities that stay in the original chunk. For example, in the second Activision chunk, there are two entities ``Call of Duty'' and ``Candy Crush'' in the \texttt{Product} category, where the former stayed in its chunk as the paired IBM chunk only had one entity under the \texttt{Product} category. This issue can be partially mitigated if we allow one vs many pairing for the matching. Furthermore, named entity counts may be incorporated as an external covariate for the clustering algorithm to match chunks with a similar number of entities.

Overall, while some implementational issues remain to be solved to further enhance the anonymization performance, it has been clearly demonstrated using this empirical example that our named entity swapping results in a significant decrease in the metadata discovery rate by the state-of-the-art LLMs. A detailed inspection of the swapped chunks has shown that part of the success stems from the fact that swapping breaks some of the combinations of named entities that may act as quasi-identifiers by moving part of these combinations over to the other chunk or introducing another quasi-identifiers from the paired chunk to make it difficult to associate post-swap chunk to a single source. Importantly, this is all done automatically without any specific instructions as to what particular combinations of terms are risky for each document. Hence, it is particularly suited for anonymization tasks involving a large corpus of texts with too many metadata categories to prevent from disclosure by providing such specific instructions to each document.
\section{Discussion \& Conclusion}\label{sec:DiscConcl}
In this work, we introduced an anonymization workflow, which can be used for the prevention of metadata disclosure from a corpus of documents sharing structural similarities. The specific method we propose is called named entity swapping, which combines named entity recognition and a procedure in statistical disclosure control called data swapping. This method achieves anonymization through interchanging some named entities that may potentially act as quasi-identifiers between a pair of text chunks. To establish an overall balance in the reduction of risk and utility for the post-swap corpus as a result of this operation, we also introduced a decision-theoretic formulation of the named entity swapping, which extends the prior work done by \cite{GomatamDecision2005} for categorical data. One major extension we made in this part is the introduction of a clustering-based weighting scheme on the text embedding space via the spherical clustering algorithm in \texttt{speroids} package. The resulting anonymization scheme is highly customizable and can be readily extended to incorporate recent advances in text embeddings. We demonstrated in the application involving earnings call transcripts that the proposed methodology is indeed able to anonymize some chunks to an extent that recent state-of-the-art LLMs are no longer able to guess the correct company names.

Although the application in Section \ref{sec:Application} is solely based on the anonymization of earnings call transcripts, the overall approach has a more general scope. As long as the corpus of documents has structural similarities across documents and touching upon similar topics, the named entity swapping will effectively interchange key terms across documents to prevent a certain combination of entities from acting as a quasi-identifier. Other examples of text data amenable to this procedure are: medical transcripts recorded across hospitals, patent filings across companies in a specific sector, etc. These post-swap anonymized text chunks will then be able to be supplied to LLMs for further use in various predictive tasks.

The current limitation of this approach is that the swapping can only be performed for a small fraction of the text chunks present in the data. However, the decision-theoretic framework for swapping can be naturally extended to allow for swapping to run several times with a change in constraints $\mathcal{C}$ to prevent swaps that have been performed before from taking place. Furthermore, other information suppressive measures, such as summarization, can be performed on text chunks that are not suitable for swapping.

\bibliography{references}
\begin{appendices}
{
\titlespacing*{\section}{0pt}{0pt}{0pt}
\section{LLM Specifications and the Prompt}\label{appendix:LLM_Prompt}
}
For predictive checks in Section \ref{subseq:PredictiveChecks}, two state-of-the-art LLMs are used. Below are the detailed specifications of these two models and the prompts given to them:
\begin{itemize}
    \item LLM Model: Google Gemini 2.5 Flash (04-17 Preview)
    \begin{itemize}
        \item Temperature: 0.0
        \item Knowledge Cutoff: January 2025
        \item Top P: 0.95
        \item Top K: 64
    \end{itemize}
    \item LLM Model: OpenAI GPT-4o (2024-11-20)
    \begin{itemize}
        \item Temperature: 0.0
        \item Knowledge Cutoff: Oct 01, 2023 
        \item Top P: 1.0
        \item Top K: 50
    \end{itemize}
\end{itemize}
The prompt given to these LLMs for the company name prediction task is as follows:
\begin{quote}
    You are tasked to guess the company name of the following earnings call transcript using the content of the text. Reply ONLY with the ticker symbol (in Capital Letters) of your best guess. DO NOT output anything other than that.

    [earnings call text chunk].
\end{quote}

{
\titlespacing*{\section}{0pt}{0pt}{0pt}
\section{Earnings Call Data}\label{appendix:Data}
}
Below are the names and dates of 35 information sector companies (as classified by NAICS) based on earnings calls held in Q1 2018:
\vspace{-0.3cm}
\begin{table}[ht]
    \centering
    \begin{tabular}{|l|l|}
\hline
\textbf{Company} & \textbf{Date} \\
\hline
ATT Inc. & Jan 31, 2018 \\
Activision Blizzard, Inc. & Feb 08, 2018 \\
Adobe Systems Incorporated & Mar 15, 2018 \\
Akamai Technologies, Inc. & Feb 06, 2018 \\
Ansys, Inc. & Feb 22, 2018 \\
Autodesk, Inc. & Mar 06, 2018 \\
Automatic Data Processing, Inc. & Jan 31, 2018 \\
Cadence Design Systems, Inc. & Jan 31, 2018 \\
Charter Communications, Inc. & Feb 02, 2018 \\
CoStar Group, Inc. & Feb 22, 2018 \\
Electronic Arts Inc. & Jan 30, 2018 \\
Etsy, Inc. & Feb 27, 2018 \\
FactSet Research Systems Inc. & Mar 27, 2018 \\
Fidelity National Information Services, Inc. & Feb 06, 2018 \\
Fortinet, Inc. & Feb 05, 2018 \\
Global Payments Inc. & Feb 15, 2018 \\
HCP, Inc. & Feb 13, 2018 \\
International Business Machines Corporation & Jan 18, 2018 \\
Intuit Inc. & Feb 22, 2018 \\
Microsoft Corporation & Jan 31, 2018 \\
Netflix, Inc. & Jan 22, 2018 \\
News Corporation & Feb 08, 2018 \\
Oracle Corporation & Mar 19, 2018 \\
PTC Inc. & Jan 17, 2018 \\
Palo Alto Networks, Inc. & Feb 26, 2018 \\
Paycom Software, Inc. & Feb 06, 2018 \\
Roper Technologies, Inc. & Feb 02, 2018 \\
Salesforce.com, inc. & Feb 28, 2018 \\
ServiceNow, Inc. & Jan 31, 2018 \\
Synopsys, Inc. & Feb 21, 2018 \\
T-Mobile US, Inc. & Feb 08, 2018 \\
Take-Two Interactive Software, Inc. & Feb 07, 2018 \\
The Walt Disney Company & Feb 06, 2018 \\
VeriSign, Inc. & Feb 08, 2018 \\
Verizon Communications Inc. & Jan 23, 2018 \\
\hline
\end{tabular}
    \caption{All 35 Firms in the Information Sector Q1 2018 Earnings Call Data Set}
    \label{tab:my_label}
\end{table}

Transcripts for each earnings call presentation were obtained from S\&P Capital IQ platform accessed on April 6th, 2025. 
\newpage
{
\titlespacing*{\section}{-1.3pt}{-1.3pt}{-1.3pt}
\section{Figures 4 and 5}\label{appendix:Figures4and5}
}
\begin{figure*}[h]
\begin{center} 
    \begin{subfigure}[t]{\textwidth}
        \includegraphics[height=3.2cm,width=\linewidth]{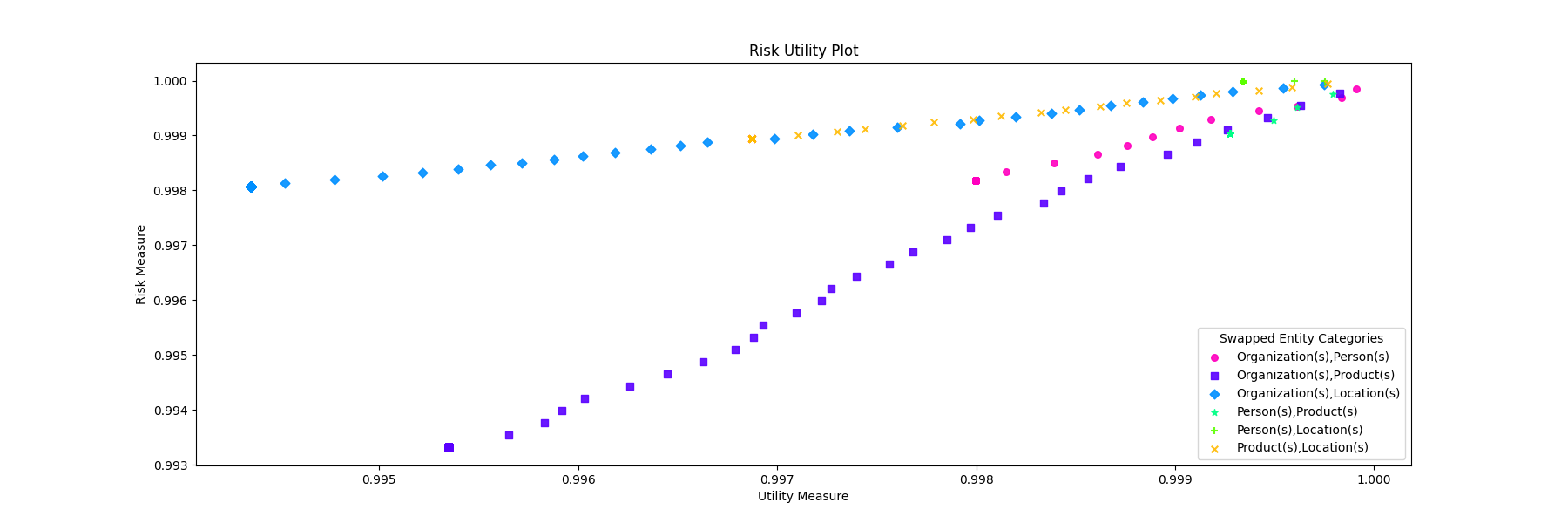}
        \caption{$N=1.0\times 10^{20}$, $f=$ PKB, $K=10$ and $\epsilon = 1.0\times 10^{-10}$}
    \end{subfigure}
    \begin{subfigure}[t]{\textwidth}
        \includegraphics[height=3.2cm,width=\linewidth]{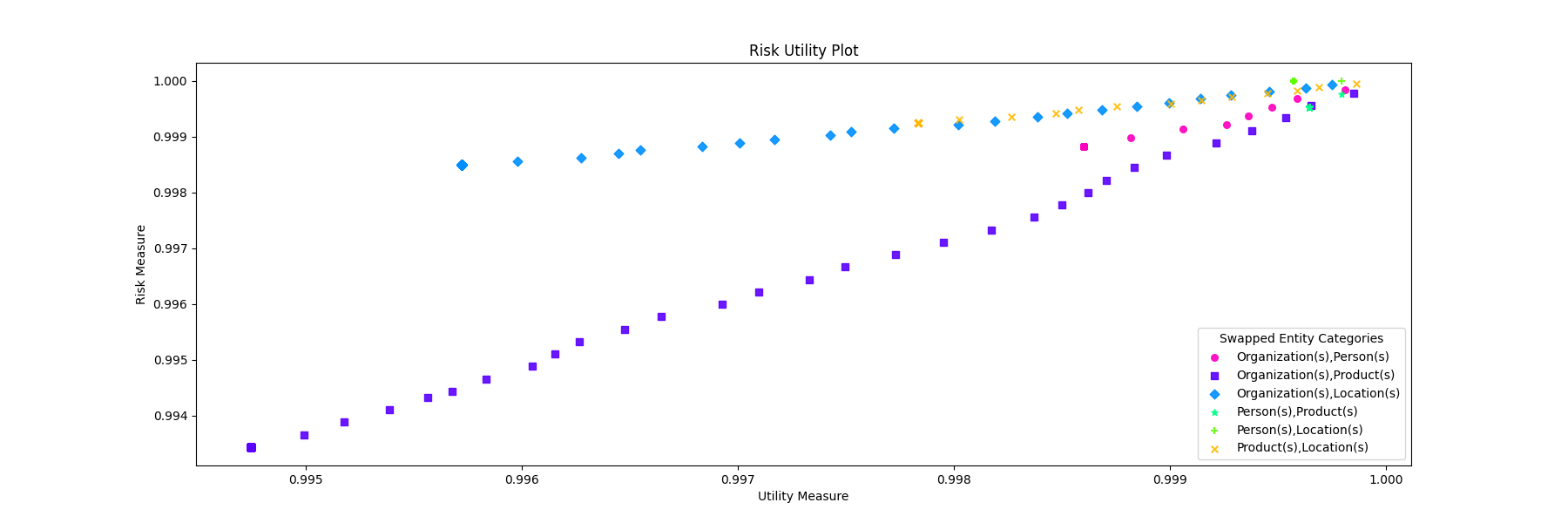}
        \caption{$N=1.0\times 10^{20}$,$f=$ PKB, $K=30$ and $\epsilon = 1.0\times 10^{-10}$}
    \end{subfigure}    

    \begin{subfigure}[t]{\textwidth}
        \includegraphics[height=3.2cm,width=\linewidth]{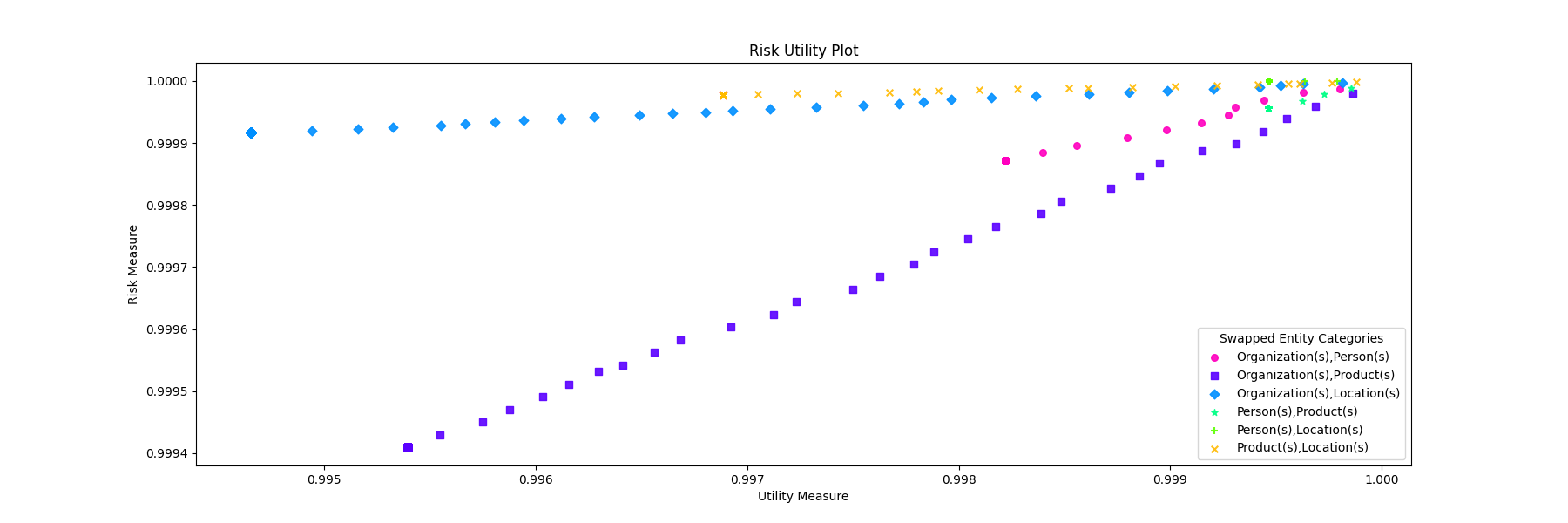}
        \caption{$N=1.0\times 10^{30}$,$f=$ PKB, $K=10$ and $\epsilon = 1.0\times 10^{-10}$}
    \end{subfigure}
    
    \begin{subfigure}[t]{\textwidth}
        \includegraphics[height=3.2cm,width=\linewidth]{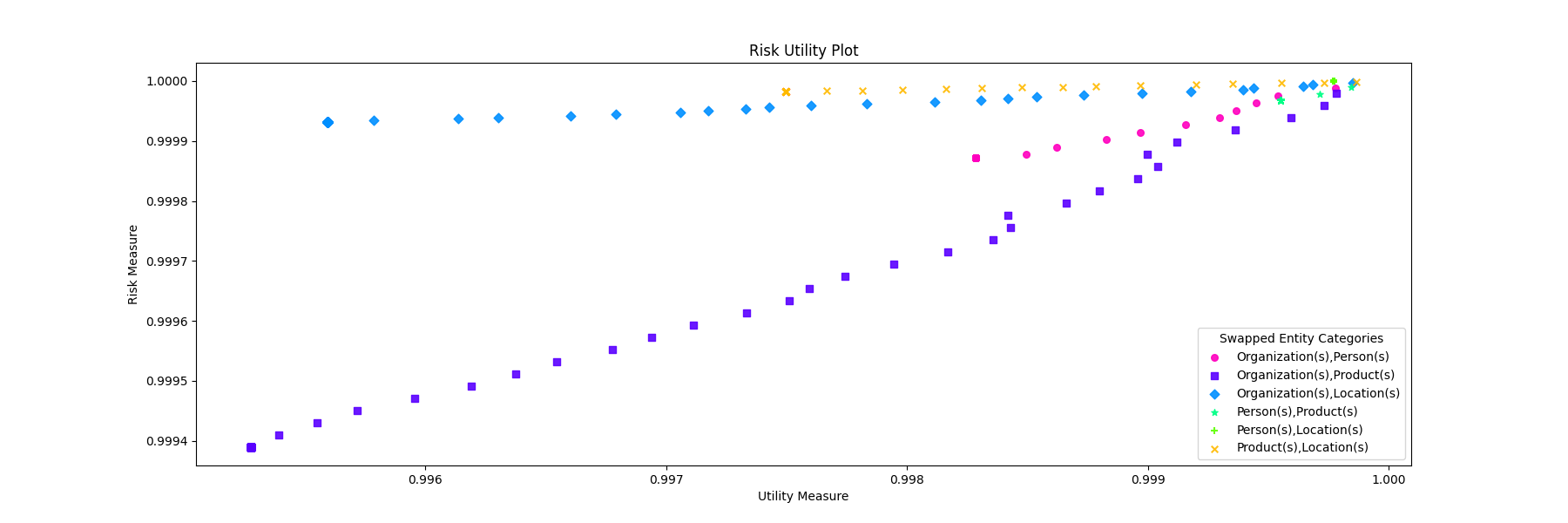}
        \caption{$N=1.0\times 10^{30}$,$f=$ PKB, $K=30$ and $\epsilon = 1.0\times 10^{-3}$}
    \end{subfigure}
    
    \begin{subfigure}[t]{\textwidth}
        \includegraphics[height=3.2cm,width=\linewidth]{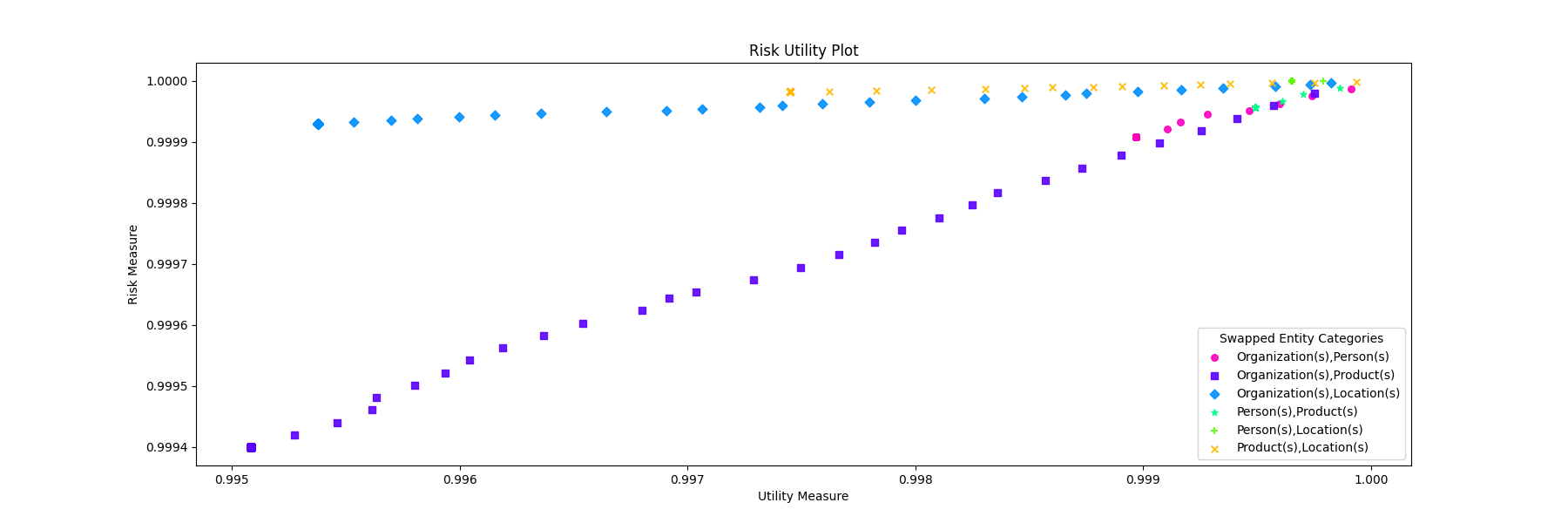}
        \caption{$N=1.0\times 10^{30}$,$f=$ PKB, $K=30$ and $\epsilon = 1.0\times 10^{-10}$}
    \end{subfigure}
    \caption{Risk-Utility Plots for the Rest of the PKB Mixtures}
    \label{fig:PKBrest}
\end{center}
\end{figure*}
\begin{figure*}
\begin{center}
    \begin{subfigure}[t]{\textwidth}
        \includegraphics[height=3.2cm,width=\linewidth]{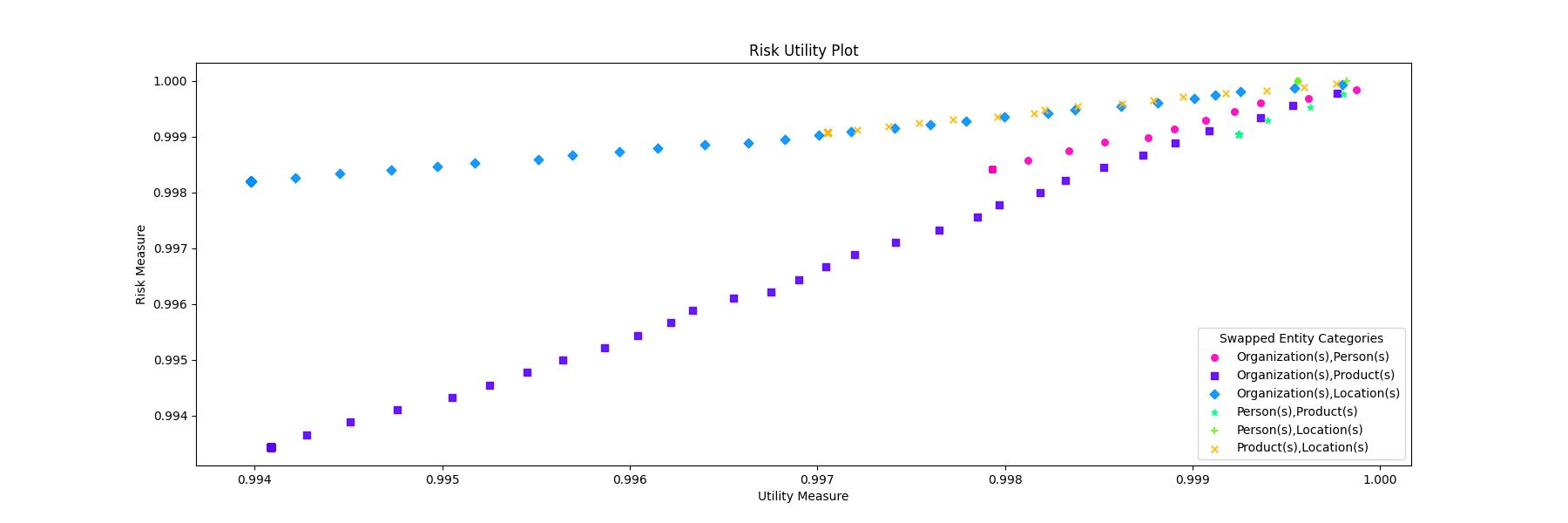}
        \caption{$N=1.0\times 10^{20}$, $f=$ sCauchy, $K=10$ and $\epsilon = 1.0\times 10^{-3}$}
    \end{subfigure}
    
    \begin{subfigure}[t]{\textwidth}
        \includegraphics[height=3.2cm,width=\linewidth]{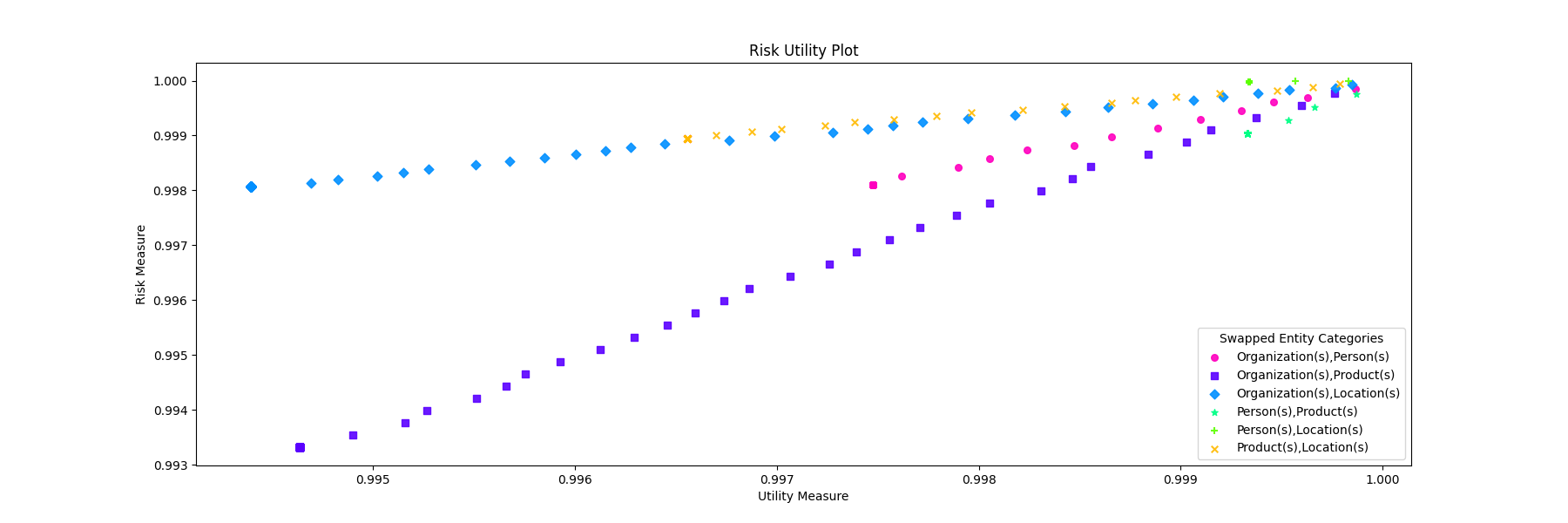}
        \caption{$N=1.0\times 10^{20}$, $f=$ sCauchy, $K=10$ and $\epsilon = 1.0\times 10^{-10}$}
    \end{subfigure}    

    \begin{subfigure}[t]{\textwidth}
        \includegraphics[height=3.2cm,width=\linewidth]{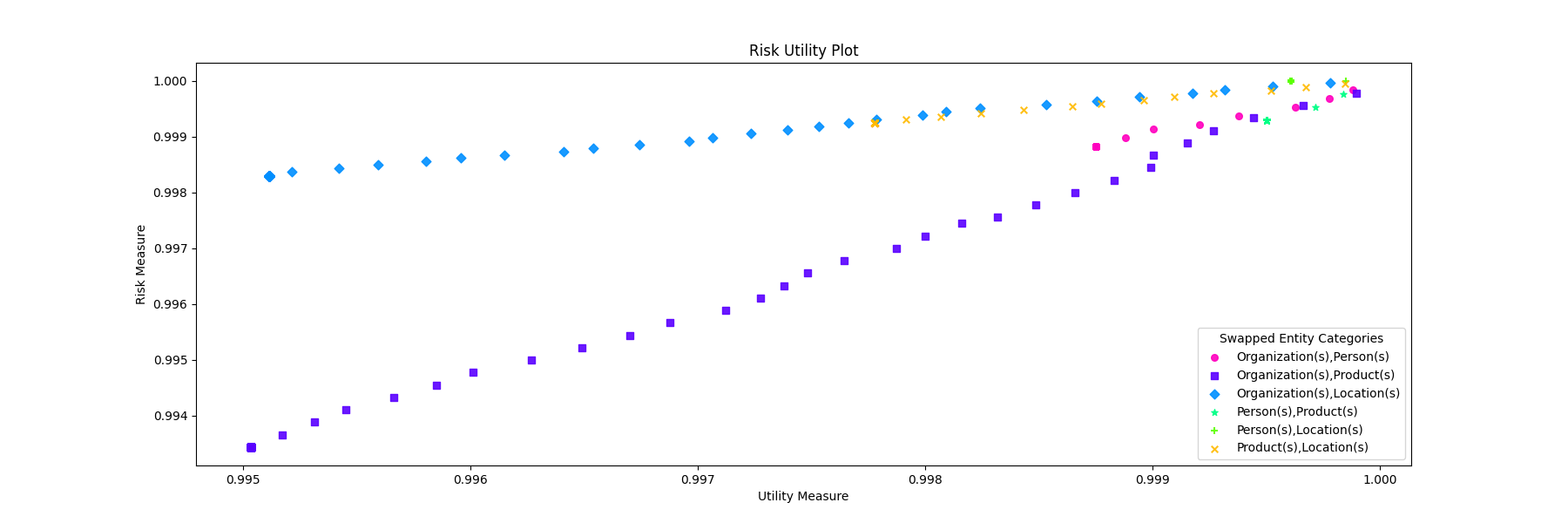}
        \caption{$N=1.0\times 10^{20}$, $f=$ sCauchy, $K=30$ and $\epsilon = 1.0\times 10^{-10}$}
    \end{subfigure}
    
    \begin{subfigure}[t]{\textwidth}
        \includegraphics[height=3.2cm,width=\linewidth]{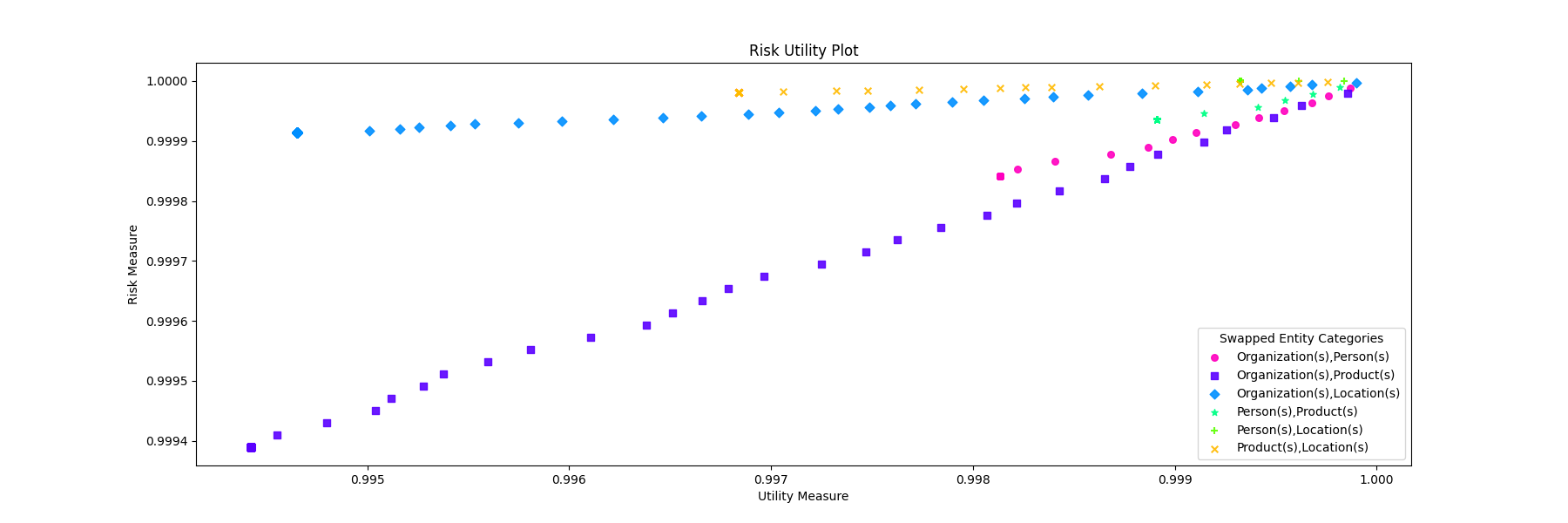}
        \caption{$N=1.0\times 10^{30}$, $f=$ sCauchy, $K=10$ and $\epsilon = 1.0\times 10^{-3}$}
    \end{subfigure}
    
    \begin{subfigure}[t]{\textwidth}
        \includegraphics[height=3.2cm,width=\linewidth]{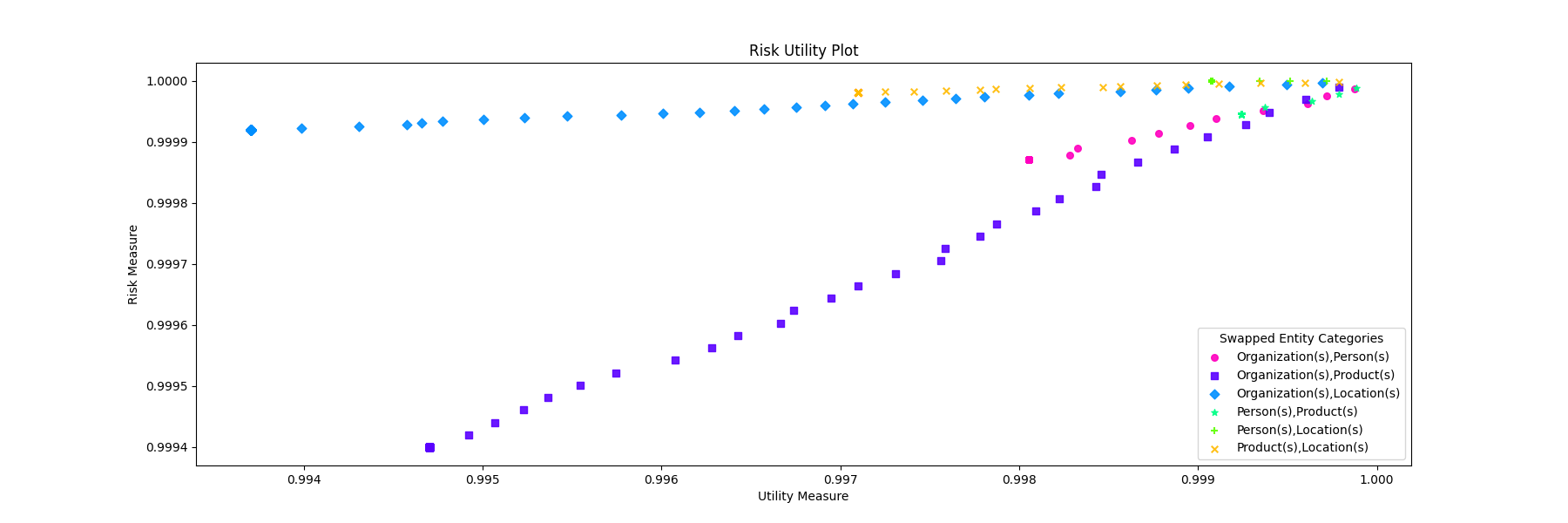}
        \caption{$N=1.0\times 10^{30}$, $f=$ sCauchy, $K=10$ and $\epsilon = 1.0\times 10^{-10}$}
    \end{subfigure}
    \caption{Risk-Utility Plots for the Rest of the Spherical Cauchy Mixtures}
    \label{fig:sCrest}
\end{center}
\end{figure*}

\end{appendices}


%
%
%
%

\end{document}